\documentclass[10pt,aps,prl,unsortedaddress,showpacs,floatfix,nofootinbib,twocolumn,longbibliography]{revtex4-1}
\usepackage[colorlinks=true,breaklinks=false]{hyperref}

\usepackage{graphicx}
\usepackage[normalem]{ulem}
\usepackage{amsmath,amstext,amssymb}
\usepackage{dcolumn}
\usepackage{bm}
\usepackage[english]{babel}
\usepackage{xcolor}

\newcommand{\mic}{~{\rm \mu m}}

\newcommand{\dshg}{d_{\rm SHG}^{\rm GVM}}
\newcommand{\ddfg}{d_{\rm DFG}^{\rm GVM}}
\newcommand{\kqpm}{k_\Lambda}
\newcommand{\wrr}{\omega_{\rm RR}}
\begin{document}

\title{Parametrically Tunable Soliton-Induced Resonant Radiation by Three-Wave Mixing}

\author{B.B. Zhou,$^1$ X. Liu,$^1$ H.R. Guo,$^{1,2}$ X. L. Zeng,$^{1,3}$ X. F. Chen,$^4$ H. P. Chung,$^5$ Y. H. Chen,$^5$ and M. Bache$^{1,*}$}

\affiliation{
$^1$DTU Fotonik, 
Technical University of Denmark,
DK-2800 Kgs. Lyngby, Denmark\\
$^2$Present address: {\'E}cole Polytechnique F{\'e}d{\'e}rale de Lausanne, CH-1015 Lausanne, Switzerland\\
$^3$Permanent address: Key Laboratory of Special Fiber Optics and Optical Access Networks, Shanghai University, Shanghai 200072, China\\
$^4$Department of Physics and Astronomy, Shanghai Jiao Tong University,
Shanghai 200240, China \\
$^5$Department of Optics and Photonics, National Central University, Jhongli 320, Taiwan\\
$^*$Corresponding author: moba@fotonik.dtu.dk\\
\today }

\begin{abstract}
\noindent
We show that a temporal soliton can induce resonant radiation by three-wave mixing nonlinearities.
This constitutes a new class of resonant radiation whose spectral positions  are parametrically tunable.
The experimental verification is done in a periodically poled lithium niobate crystal, where a femtosecond near-IR soliton is excited and resonant radiation waves are observed exactly at the calculated soliton phase-matching wavelengths via the sum- and difference-frequency generation nonlinearities.
This extends the supercontinuum bandwidth well into the mid-IR to span 550-5000 nm and the mid-IR edge is parametrically tunable over 1000 nm by changing the three-wave mixing phase-matching condition. The results are important for bright and broadband supercontinuum generation and for frequency comb generation in quadratic nonlinear microresonators.


\end{abstract}


\maketitle


\noindent
The temporal optical soliton is quite remarkable: While often desired because it retains its form despite dispersive effects \cite{mollenauer:1980}, perturbing the perfect solitary shape may also lead to phase-matching of a so-called resonant radiation (RR) wave \cite{Wai:1986} (also known as soliton-induced optical Cherenkov radiation \cite{akhmediev:1995}).
RR waves are today considered a coherent source of laser radiation \cite{Skryabin:2010}, in particular for supercontinuum generation \cite{dudley:2006} where they contribute coherently for extending  the supercontinuum bandwidth, as well as in ultrashort pulse generation in the UV and the mid-IR \cite{joly:2011,Mak2013,Belli:2015-VUV-DW,Ermolov-2015-PhysRevA.92.033821}.
Traditionally RR waves are generated by four-wave mixing (4WM) through the Kerr self-phase modulation (SPM) term $|A|^2A$ \cite{beaud:1987,Wise:1988,Gouveia-Neto:1988}. Recently other non-standard 4WM terms were shown to support RR waves, namely the conjugate SPM term $|A|^2A^*$ \cite{rubino:2012,petev.PhysRevLett.111.043902,conforti-PhysRevA.88.013829-2013,Rubino:2012-SciRep} and third-harmonic generation term $A^3$ \cite{Conforti2013,loures-PhysRevLett.115.193904}.

\begin{figure}[tb]
\centerline{\includegraphics[width=\linewidth]
{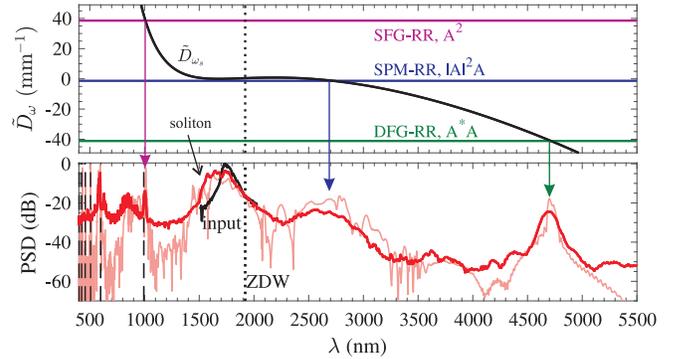}
}
\caption{Supercontinuum recorded for $\lambda_0=1.75\mic$ and $I_0=150 ~{\rm GW/cm^2}$, using a 10 mm PPLN with $\Lambda=30.0\mic$; the experimental data (thick red) are directly compared to a numerical simulation (thin light red) and calculated higher-order quasi-phase matching resonances (dashed lines). The power-spectral densities (PSDs) are normalized to the peak input PSD, and so that the average output power matches the input power. The top plot shows the theoretical RR phase-matching conditions to the soliton using $\lambda_s=1.68\mic$. ZDW: zero-dispersion wavelength ($1.92\mic$).
}
\label{fig:PSD_PM}
\end{figure}

Phase-mismatched (cascaded) three-wave mixing (3WM) in quadratic nonlinear crystals can generate a negative self-defocusing Kerr-like nonlinearity \cite{desalvo:1992,liu:1999}, and when a temporal soliton is excited \cite{ashihara:2002} this gives octave-spanning supercontinua \cite{Langrock:2007,Phillips:2011-ol,Phillips:2011,Guo-LN-exp-2015} that are filament free \cite{zhou:2012,Zhou:2015,Zhou:2015-OL,Zhou2016-aplp}. So far, only RR waves generated by the cascaded self-defocusing SPM effect have been verified \cite{Zhou:2015,Zhou:2015-OL,Zhou2016-aplp}. In this Letter we show that a new class of RR waves exist in quadratic nonlinear crystals, induced by the soliton through the 3WM processes sum-frequency generation (SFG, $A^2$) and difference-frequency generation (DFG, $A^*A$). These RR waves contribute coherently to the supercontinuum, making it brighter and more broadband, and remarkably their center wavelengths are parametrically tunable by adjusting \textit{only} the 3WM phase-matching conditions, giving an additional control over the supercontinuum often lacking in the 4WM RR case.
Here we show direct experimental proof of 3WM RR waves 
excited by an IR soliton in a periodically poled lithium niobate (PPLN) crystal (Fig. \ref{fig:PSD_PM}). The SFG and DFG RR waves are observed, and we demonstrate how their center wavelengths can be strongly tuned by changing the 3WM phase-matching conditions. The DFG RR wave in particular provides a tunable extension of the supercontinuum well into the mid-IR ($4.0$-$5.5\mic$ range, important for ultrafast vibrational spectroscopy).

In order to describe these novel RR waves the nonlinear terms cannot be truncated, which implies resolving the electric field on a carrier level. We use the nonlinear analytic envelope equation \cite{conforti:2010PRA,Conforti2013}, whose formalism conveniently retains an envelope-like equation. The $\chi^{(2)}$-$\chi^{(3)}$ nonlinear dynamics at the pump frequency $\omega_1$ is described in a single equation of the $e$-polarized pump envelope $A$ in the co-moving reference frame $(\zeta,\tau)$ \cite{bache:2016-NAEE,Baronio:2012,PRL-2017-suppl}
\begin{multline}\label{eq:NAEE}
i\partial_\zeta A+\hat D_{\omega_1} A=
\\
-\kappa^{(2)}
[
\tfrac{1}{2}A^2 e^{-i\omega_1 \tau-i\Delta _{\rm pg}\zeta}
+A^*A e^{i\omega_1 \tau+i\Delta _{\rm pg}\zeta}
]_+
-\kappa^{(3)}
[|A|^2A
\\
+|A|^2A^* e^{i2\omega_1\tau + i2\Delta _{\rm pg} \zeta}+\tfrac{1}{3}A^3 e^{-i2\omega_1\tau - i2\Delta _{\rm pg} \zeta}
]_+
\end{multline}
Self-steepening terms and delayed Raman effects \cite{bache:2016-NAEE} are here neglected as they do not influence the following phase-matching analysis. The nonlinear parameters are $\kappa^{(2)}\propto \chi^{(2)}$ and $\kappa^{(3)}\propto \chi^{(3)}$, related to quadratic and cubic nonlinear terms, respectively. The operator $\hat D_{\omega_1}=\sum_{m=2}m!^{-1}k_m(\omega_1)(i\partial_\tau)^m$ accounts for dispersion in time domain, where $k_m(\omega)=d^mk(\omega)/d\omega^m$ are the higher-order dispersion coefficients. The dispersion is conveniently evaluated exactly in frequency domain as $\tilde D_{\omega_1}(\omega)=k(\omega)-k_1(\omega_1)(\omega-\omega_1)-k(\omega_1)$.
The constant term $\Delta_{\rm pg}=\omega_1 k_1(\omega_1)-k(\omega_1)=\omega_1(1/v_g-1/v_p)$ accounts for the phase-group-velocity mismatch (carrier-envelope phase slip), where $v_g=1/k_1(\omega_1)$ is the pump group velocity and $v_p=c/n(\omega_1)$ is the pump phase velocity. Finally, the $+$ sign implies that only the positive frequency content of the nonlinear term is used \cite{conforti:2010PRA}. 

The equations will support a number of RR phase-matching conditions between a soliton at frequency $\omega_s$ (without loss of generality we can take $\omega_s=\omega_1$) and a "dispersive" (i.e., non-solitonic) RR wave
\begin{align}
\label{eq:PM-SPM-moving}
\tilde D_{\omega_s}(\wrr)&=q_s,  &(\text{SPM-RR}, \; |A|^2A)
\\
\label{eq:PM-cSPM-moving}
\tilde D_{\omega_s}(\wrr)&=-q_s+2\Delta_{\rm pg},  &(\text{cSPM-RR}, \; {|A|}^2A^*)
\\
\label{eq:PM-THG-moving}
\tilde D_{\omega_s}(\wrr)&=3q_s-2\Delta_{\rm pg},  &(\text{THG-RR}, \; {A^3})
\\
\label{eq:PM-SFG-moving}
\tilde D_{\omega_s}(\wrr)&=2q_s-\Delta_{\rm pg},  &(\text{SFG-RR}, \; {A^2})
\\
\label{eq:PM-DFG-moving}
\tilde D_{\omega_s}(\wrr)&=\Delta_{\rm pg},  &(\text{DFG-RR}, \; {A^*A})
\end{align}
These were found by inserting the ansatz  $A(\zeta,\tau)=F_s(\tau)e^{i q_s\zeta}+g(\zeta,\tau)$ \cite{skryabin:2005} into Eq. (\ref{eq:NAEE}) \cite{PRL-2017-suppl}. Here, $F_s(\tau)e^{i q_s\zeta}$ is the exact nonlinear solution when taking into account only the SPM nonlinearity and GVD, $F_s$ is the soliton envelope, $q_s$ is the soliton nonlinear wavenumber  
and $g$ is the RR wave \cite{PRL-2017-suppl}.
The first phase-matching condition is the traditional RR induced by SPM, the second is the RR from the "conjugate SPM" term (a.k.a. negative-frequency RR), while the third is the "third-harmonic generation RR" or simply THG-RR.
The last two are the new 3WM phase-matching conditions from the SFG and DFG $\chi^{(2)}$ nonlinear terms, here presented for the first time.
In quadratic nonlinear crystals, the Kerr-like SPM-RR were predicted \cite{bache:2010e} and experimentally confirmed \cite{Zhou:2015,Zhou:2015-OL}, and the Kerr-like cSPM-RR was studied numerically \cite{conforti-PhysRevA.88.013829-2013}.



For a physical interpretation it is instructive to transform the interaction back to the lab-frame coordinate by using the soliton dispersion relation $k_s(\omega)=k(\omega_s)+(\omega-\omega_s)/v_{g,s}+q_s$. 
In the SPM-RR case we immediately get the well-known dispersion relation $k(\wrr)= k_s(\wrr)$ \cite{skryabin:2005}, caused by self-acting 4WM (i.e. "degenerate" \cite{skryabin:2005} 4WM).
In the SFG case (the $A^2$ term), the rewritten phase-matching condition is 
\begin{eqnarray}
k(\wrr)&=2k_{s}(\wrr)-\wrr/v_{g,s}
=k_s(\omega_a)+k_s(\omega_b)
\label{eq:SFG}
\end{eqnarray}
i.e., SFG between soliton photons at two different frequencies, and where energy conservation dictates $\omega_a+\omega_b=\wrr$. If we take $\omega_a=\omega_s$ then $\omega_b=\wrr-\omega_s$. We now show how the second-harmonic generation (SHG) phase mismatch parameter $\Delta k _{\rm SHG}$ is affecting this phase-matching condition: By expanding the dispersion operator in Eq. (\ref{eq:PM-SFG-moving}) around the second-harmonic (SH) frequency of the soliton frequency $\omega_2=2\omega_s$ we get
\begin{eqnarray}
\tilde D_{2\omega_s}(\wrr) -(\wrr-2\omega_s)\dshg+\Delta k_{\rm SHG}=2q_s
\label{eq:SH-nonlocal}
\end{eqnarray}
where $\dshg=k_1(\omega_s)-k_1(2\omega_s)$ is the group-velocity mismatch coefficient between the soliton and its SH.  Clearly, the SHG phase-mismatch parameter $\Delta k_{\rm SHG}=k(2\omega_s)-2k(\omega_s)$ allows for a tunable control over the spectral position of the RR wave. In a birefingent critically phase-mismatched interaction $\Delta k_{\rm SHG}$ can be tuned by the crystal angle, while in a non-critical phase-mismatched interaction (e.g., PPLN) $\Delta k_{\rm SHG}$ is effectively controlled by the poling pitch.
Interestingly, Eq. (\ref{eq:SH-nonlocal}) is identical to the SH nonlocal resonance condition \cite{bache:2007a,bache:2008} that recently was confirmed  experimentally \cite{zhou:PhysRevA.90.013823}. Thus, in the defocusing soliton parameter range, the tunable nonlocal SH wave in \cite{zhou:PhysRevA.90.013823} is identical in nature to the SFG-RR wave we predict here. Additionally, the simulations in \cite{conforti:2010PRA} seem also to show an SFG-RR wave. However, we stress that these previous observations did not identify these peaks as resonant radiation.

The DFG case can similarly be written as
\begin{eqnarray}
k(\wrr)&=\wrr/v_{g,s}
=k_s(\omega_a)+k_s(\omega_b)
\label{eq:DFG}
\end{eqnarray}
where 
$\omega_a-\omega_b=\wrr$. Thus, the physics behind this condition is DFG between soliton photons at different frequencies, and it explains why the soliton wavenumber $q_s$ is absent in Eq. (\ref{eq:PM-DFG-moving}). Expanding this phase-matching condition around the DFG frequency reveals a relation similar to the SFG case, namely that the DFG phase-mismatch can tune the RR spectral position. 
The DFG-RR phase-matching condition is not easy to fulfill. 
To see that, Eq. (\ref{eq:PM-DFG-moving}) is expressed 
as
\begin{align}\label{eq:PM-DFG-v}
v_{\rm ph}(\wrr)=v_{g,s}
\end{align}
i.e., that the RR phase velocity $v_{\rm ph}(\wrr)$ is the same as the soliton group velocity.
This is equivalent to the so-called \textit{velocity-matching condition}, encountered e.g. when generating THz radiation in a quadratic nonlinear crystal through DFG \cite{Nahata1996}. The intuitive explanation behind the THz velocity-matching condition is that the THz carrier wave relies on the different colors of the pump wave having the same group velocity, i.e., that they move as a single wavepacket. Essentially the same can be said about the soliton in the DFG-RR case, because due to the straight-line dispersion of the soliton, its photons will move with the same group velocity no matter at what frequency they are taken from.
Matching the phase- and group-velocities is not easy, especially because the DFG process implies that the converted photon is at a lower frequency than the pump photons. 
This makes velocity-matching virtually impossible, except when the converted wave lies beyond an IR resonance (like the THz case) where the drawback is a very low yield.

However, by exploiting quasi-phase matching (QPM) we can achieve velocity matching in the same transparency window as the soliton, and thereby observe the DFG-RR for the first time. 
Taking a square-grating periodic-poling structure of the quadratic nonlinearity with pitch $\Lambda$, the 3WM conditions change to
\begin{align}
\label{eq:PM-SFG-moving-QPM}
\tilde D_{\omega_s}(\wrr)&=2q_s-\Delta_{\rm pg}+ \kqpm, \; &(\text{SFG-RR}, \; {A^2})\\
\label{eq:PM-DFG-moving-QPM}
\tilde D_{\omega_s}(\wrr)&=\Delta_{\rm pg}-\kqpm, \; &(\text{DFG-RR}, \; {A^*A})
\end{align}
We see that by tuning the QPM wavenumber $k_\Lambda=2\pi/\Lambda$ we can now manipulate the phase-matching conditions and get tunable control over the RR frequency. 

The soliton we excite here is a bright self-defocusing temporal soliton. The negative (self-defocusing) nonlinearity is created through strongly phase-mismatched (i.e., cascaded) second-harmonic generation (SHG). Essentially the pump wave will experience a Kerr-like nonlinear refractive index $n_{2, \rm casc}\propto -(\chi^{(2)})^2/\Delta k_{\rm SHG}$ \cite{desalvo:1992}, and this will compete with the intrinsic material self-focusing Kerr nonlinearity $n_{2, \rm Kerr}\propto\chi ^{(3)}$. If the SHG phase-mismatch $\Delta k_{\rm SHG}$ is made suitably small, and the residual \textit{effective} nonlinear refractive index $n_{2, \rm eff}=n_{2, \rm casc}+n_{2, \rm Kerr}$ becomes negative, the soliton can be excited in the normal group-velocity dispersion (GVD) regime [$k_2(\omega_1)>0$] below the ZDW.

The experimental setup was similar to \cite{Zhou:2015}, and consisted of only the pump, a silver-mirror telescope, and the PPLN crystal. The pump laser was a 1 kHz OPA system and wavelengths 1.55-$1.85\mic$ were used, all located below the ZDW of LN. The pump pulse duration was 60 fs and close to transform limit, and was loosely collimated before the crystal (0.5 mm FWHM spot size). Several bulk PPLN crystals with multi-grating structures were used with pitch gratings from $\Lambda=27.0$-$31.6\mic$, all designed to exploit the large $d_{33}$ quadratic nonlinearity; in this range $|n_{2, \rm casc}|/n_{2, \rm Kerr}\simeq 1.5-2.0$. The spectrum was measured in the mid-IR ($\lambda>2.3\mic$) with an FPAS-1600 spectrometer (Infrared Systems) with a cooled MCT detector, and long-pass filters were used to selectively cover the 2-6$\mic$ range. In the visible and near-IR range compact spectrometers were used, based in Si and InGaAs CCD detectors, respectively.


Figure \ref{fig:PSD_PM} shows a typical high-intensity spectrum. The pump pulse (80 nm FWHM) has experienced massive broadening, and a supercontinuum is formed spanning over 3 octaves (550-5000 nm). The soliton has clearly broadened to the blue: a "center-of-mass" calculation gave $\lambda_s=1.68\mic$, which was then used to calculate the RR phase-matching curves from the expressions derived above. The soliton wavenumber $q_s$ was estimated to that of a $T_s=10$ fs soliton (a typical value from simulations); assuming that such a soliton will have unity soliton order one can use the expression for $q_s=n_{2,\rm eff}I_s\omega_s/(2c)$ \cite{bache:2010e} and that of the effective soliton order \cite{bache:2007} to get $q_s=-2k_2(\omega_s)/T_s^2$.
By comparing these curves with the experimental data, we identify a number of RR peaks: Firstly, the broad mid-IR peak above the ZDW is the Kerr SPM-RR wave, identical in nature to the recent observations in other crystals \cite{Zhou:2015,Zhou:2015-OL}. Secondly, a peak is located at 1002 nm. Even if this is close to a QPM phase-matching line (dashed line), we show below evidence that this is indeed the SFG-RR wave. Finally, the peak at 4700 nm is the DFG-RR wave.
In \cite{Langrock:2007,Phillips:2011-ol,Phillips:2011}, mid-IR spectral peaks were also observed, but by carrying out a similar phase-matching analysis as above we can only conclude that these were SPM-RR waves. 
In the low-wavelength range there are too many gaps to form a continuum. Many narrow lines were seen stemming from QPM higher-order resonances, e.g., the SHG QPM conditions $k(\omega)-2k(\omega/2)-m_0 k_\Lambda=0$, with $m_0$ odd. The plot also shows the result of a numerical simulation \cite{PRL-2017-suppl}, showing excellent quantitative agreement. 
Finally, we mention that the elusive cSPM-RR and THG-RR waves were not observed.

\begin{figure}[tb]
\centerline{\includegraphics[width=\linewidth]{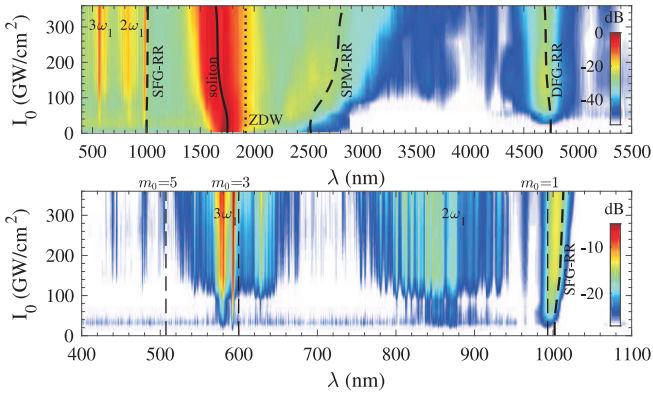}}
\caption{False-color representation of the experimental supercontinua for various intensities with the same parameters as Fig. \ref{fig:PSD_PM}. Dashed lines:  theoretical phase-matching wavelengths using the extracted soliton wavelengths (black line). Bottom: details of the visible and short-wavelength near-IR range, including calculated QPM resonances.
}
\label{fig:exp-29.5}
\end{figure}

Figure \ref{fig:exp-29.5} shows how the spectrum changes with intensity. Based on the appearance of the DWs, we estimate that the soliton forms for much lower intensities (at around 50 $\rm GW/cm^2$) than in unpoled LN \cite{zhou:2012,Zhou:2015}, which is due to the larger effective nonlinearity as QPM significantly reduces the SHG phase-mismatch. For increasing intensities the soliton becomes more blue-shifted (the black dashed line shows the calculated "center-of-mass" soliton wavelength). This is in stark contrast to the massive Raman-induced red-shift observed in unpoled LN \cite{zhou:2012,Zhou:2015}, and is a consequence of pumping close to the ZDW, which makes the soliton recoil towards the blue. The blue-shifted soliton wavelength directly affects all three phase-matching conditions. This blue-shift explains why the SPM-RR plateau red-shifts with increasing intensity. The DFG-RR phase-matched RR wave remains more or less constant, but the SFG-RR wave noticeably changes wavelength from low to high intensity as the soliton blue shifts, see bottom plot. Even if this peak lies quite close to the $m_0=1$ QPM line, there is evidence that it is indeed a DW: it clearly follows the calculated SFG-RR phase-matching as the intensity increases and it is also too broadband to be a QPM line; note in contrast how narrow the $m_0=3$ and 5 QPM lines are. For high intensities the SPM-RR plateau  flattens, and numerical simulations \cite{PRL-2017-suppl} indicate that this is due to increased pump depletion as well as significant self-steepening during the soliton formation stage. We are currently investigating this further.
The visible range contains spectral "copies" of the soliton supercontinuum at the harmonic wavelengths ($2\omega_1$, $3\omega_1$); this is due to trapped radiation caused by the cascaded (i.e. phase-mismatched) nonlinearities \cite{bache:2008,zhou:PhysRevA.90.013823,Valiulis:2011}, giving a coherent extension of the supercontinuum into the visible. The simulations indeed confirmed that the supercontinua had a high degree of coherence, also in the trapped harmonic extensions. 

\begin{figure}[tb]
\centerline{\includegraphics[width=\linewidth]{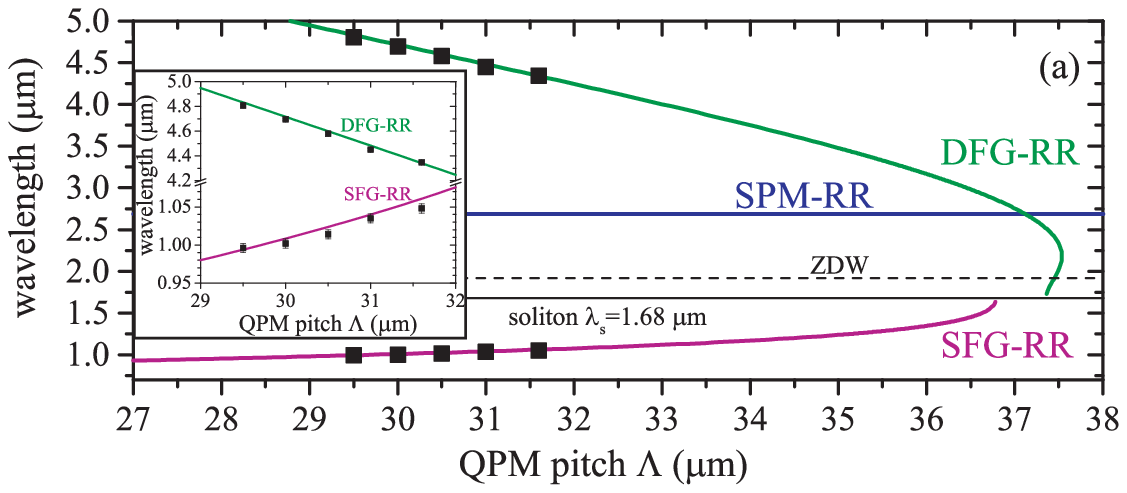}}
\centerline{\includegraphics[width=\linewidth]{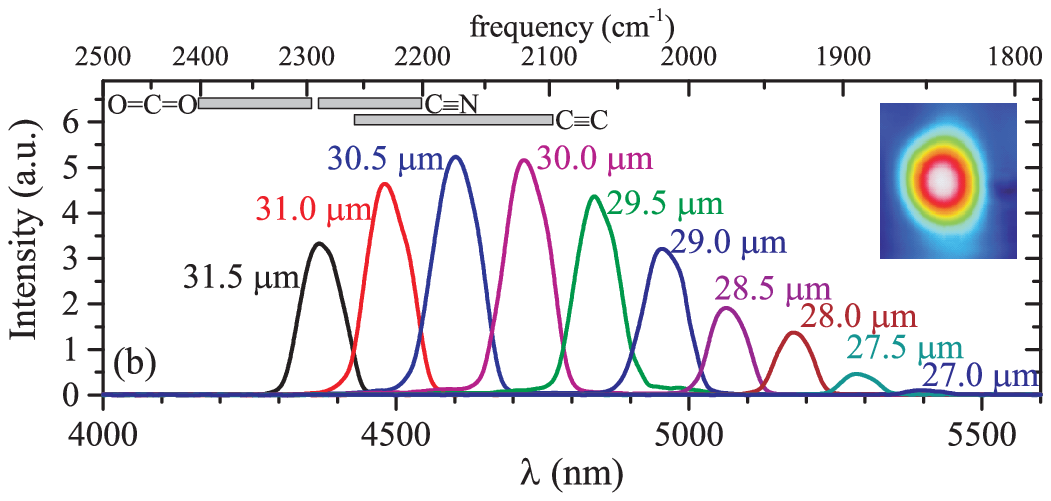}}
\caption{(a) Calculated tuning curves for SPM, DFG and SFG RR waves (using a fixed $\lambda_s=1.68\mic$), plotted with data from the same experiment as Fig. \ref{fig:PSD_PM}, but where the QPM pitch was varied. Inset: details around the experimental data. (b) Experimental mid-IR spectra showing the DFG-RR peaks, recorded with a 20 mm PPLN with 10 different QPM pitch values using $\lambda_0=1.75\mic$ and $I_0=55 ~{\rm GW/cm^2}$; note the linear y-axis. Inset: typical transverse beam profile of the long-pass filtered mid-IR RR wave.
}
\label{fig:MIR}
\end{figure}

The parametric tunability of the TWM nonlinearities gives a mid-IR edge of the supercontinuum output that is tunable, which is demonstrated in Fig. \ref{fig:MIR}(a) where the poling pitch is varied. Note also the excellent agreement with the theoretical phase-matching calculations (see zoom in the inset). 
The theoretical curves also show the further potential in tuning the 3WM RR waves: essentially the whole near- and mid-IR transparency range of LN is covered, although practically the soliton and the SPM-RR will dominate the 1.5-3.5$\mic$ range. The plot in Fig. \ref{fig:MIR}(b) confirms that indeed a broader tuning range in the mid-IR is possible: the DFG-RR waves recorded in a 20 mm long PPLN crystal with 10 different QPM pitch values. All spectra were recorded one after another so the intensity magnitudes are therefore absolute and can be related to each other. The tuning range demonstrated here has a quite specific importance: as the inset bars indicate, the IR molecular vibration absorption bands that are present in this range are IR stretching modes, which, apart from the well-known band for ${\rm CO_2}$ in the gas phase, include the important alkyne and nitrile stretching modes. This degree of tunability is quite unique, 
and it could be exploited by using an adiabatic change in the pitch along the crystal to give a broader and smoother DFG-RR peak to cover the spectral gap towards the SPM-RR.
In the 4WM case the RR position has little or no tunability except in  a gas-filled hollow-core fiber, where the pressure may control both the dispersion and the nonlinearity \cite{joly:2011,Mak2013,Belli:2015-VUV-DW,Ermolov-2015-PhysRevA.92.033821}. However, it also requires adjusting the pump power.

Similar to \cite{Zhou:2015}, we isolated the mid-IR DFG-RR waves with a long-pass filter and measured them to be sub-picosecond pulses with a significant amount of chirp (this is expected as the RR waves are inherently dispersive). 
Using a THz camera we measured the isolated mid-IR beam profiles, revealing Gaussian-like shapes (Fig. \ref{fig:MIR}(b), inset).

In conclusion we have shown soliton-induced resonant radiation mediated by $\chi^{(2)}$ three-wave mixing nonlinear terms representing sum- and difference-frequency generation ($A^2$ and $A^*A$, respectively). These provide a powerful extension of the well-known resonant radiation induced by $\chi^{(3)}$ four-wave mixing, in particular due to the broadband parametric tunability of the resonant wavelengths provided by the SFG or DFG phase-mismatch parameters. 
Our experiment was conducted in PPLN crystals, where the parametric tunability came from changing the QPM pitch. This, combined with the excitation of a self-defocusing soliton, allowed us to phase-match the DFG resonant radiation wave in the mid-IR.
The SFG and SPM resonant radiation waves were also observed and the full supercontinuum spanned over 3 octaves (550-5000 nm).
Our results could find direct use in soliton-based frequency-comb generation in microresonators \cite{DelHaye2007,Herr.2014,Brasch2016}, 
for which quadratic nonlinear materials are currently being explored for on-chip inherent harmonic conversion of the IR comb lines \cite{Jung2014,Miller2014,Ricciardi-PhysRevA.91.063839,leo-PhysRevLett.116.033901,Jung2016}. Considering that the standard cavity nonlinear model \cite{leo-PhysRevLett.116.033901} was recently adopted for the quadratic nonlinearity case \cite{Hansson2016}, which showed similar nonlinear terms as in Eq. (\ref{eq:NAEE}), we believe that the demonstrated SFG and DFG resonant radiation waves 
can provide a unique tunable control over the coherent extension of the comb lines.

\begin{acknowledgments}Support from the Danish Council for Independent Research (grant no. 11-106702) is acknowledged. X.L. Zeng acknowledges the support of National Natural Science Foundation of China (11274224) and from FP7 - Marie Curie Actions (grant no. PIIF-GA-2009–253289). M.B. acknowledges fruitful discussions with Peter Uhd Jepsen.
\end{acknowledgments}

\appendix*

\section{Appendix: Supplementary material for Parametrically Tunable Soliton-Induced Resonant Radiation by Three-Wave Mixing}

\section{The Nonlinear Analytic Envelope Equation}

In order to observe the novel resonant radiation (RR) waves, one cannot use the standard nonlinear Schr{\"o}dinger equation as it only contains the SPM term $|A|^2A$. Instead, it is necessary to model the full electrical field either by a forward-Maxwell equation approach \cite{conforti:2010,Guo:2013} or by using the so-called nonlinear analytic envelope equation (NAEE) \cite{conforti:2010PRA}. In particular the latter has the advantage of modelling carrier-wave resolved dynamics while still keeping the envelope-like equations. Specifically since in PPLN the coupling to the $o$-polarized component is zero as the crystal is cut for $\theta=\pi/2$, we can model the $\chi^{(2)}$ dynamics in the $e$-polarized pump at frequency $\omega_1$ in a single equation in the moving reference frame $\zeta=z$ and $\tau=t-z\beta_1$ \cite{bache:2016-NAEE}
\begin{widetext}
\begin{multline}\label{eq:NLSE-NAEE-Raman}
i\frac{\partial A}{\partial \zeta}+\hat D_\tau A
+\kappa^{(2)}\hat S_\tau \left[
\tfrac{1}{2}A^2 e^{-i\omega_1 \tau-i\Delta _{\rm pg}\zeta}
+|A|^2 e^{i\omega_1 \tau+i\Delta _{\rm pg}\zeta}
\right]_+
\\
+\kappa^{(3)}\hat S_\tau \Big[
(1-f_R)\left(|A|^2A+|A|^2A^* e^{i2\omega_1\tau + i2\Delta _{\rm pg} \zeta}+ \tfrac{1}{3}A^3 e^{-i2\omega_1\tau - i2\Delta _{\rm pg} \zeta}\right)
\\
+f_R \Big\{
\tfrac{1}{2}A(\zeta,\tau)
\int_{-\infty}^\infty d\tau'
h_R(\tau-\tau') e^{i2\omega_1 \tau'+i2\Delta _{\rm pg}\zeta} A^{*2}(\zeta,\tau')
\\
+\left(A(\zeta,\tau)+A^*(\zeta,\tau)e^{i2\omega_1 \tau+i2\Delta _{\rm pg}\zeta}\right)
\int_{-\infty}^\infty d\tau'
h_R(\tau-\tau')\left(\tfrac{1}{2}A^2(\zeta,\tau')e^{-i2\omega_1 \tau'-i2\Delta _{\rm pg}\zeta}+|A(\zeta,\tau')|^2\right)
\Big\}
\Big]_+=0
\end{multline}
\end{widetext}
where for notational reasons we have suppressed the dependence of $A$ on $\zeta$ and $\tau$ except in the Raman part where it is spelled out for clarity.
The nonlinear terms are
\begin{align}\label{eq:chi2}
  \kappa^{(2)}&=\frac{\omega_1\chi^{(2)}}{2n(\omega_1)c} \\
  \label{eq:chi3}
  \kappa^{(3)}&=\frac{3\omega_1\chi^{(3)}}{8n(\omega_1)c}
\end{align}
and $\hat S_\tau=1+i\omega_1^{-1}\tfrac{\partial}{\partial \tau} $ is the self-steepening operator. $f_R$ is per usual the fraction of Raman nonlinearity and $h_R(t)$ the normalized Raman response function.

Neglecting Raman, we arrive at the more simple form
\begin{align}\label{eq:NAEE-suppl}
i\partial_\zeta A&+\hat D_{\omega_1} A
+\kappa^{(2)}\left(1+i\omega_1^{-1}\partial _\tau\right)
[
\tfrac{1}{2}A^2 e^{-i\omega_1 \tau-i\Delta _{\rm pg}\zeta}
\nonumber\\
&+A^*A e^{i\omega_1 \tau+i\Delta _{\rm pg}\zeta}
]_+
\nonumber\\
&
+\kappa^{(3)}\left(1+i\omega_1^{-1}\partial _\tau\right)
[|A|^2A
\nonumber\\
&+|A|^2A^* e^{i2\omega_1\tau + i2\Delta _{\rm pg} \zeta}+\tfrac{1}{3}A^3 e^{-i2\omega_1\tau - i2\Delta _{\rm pg} \zeta}
]_+
=0
\end{align}
which once self-steepening is removed reverts to the equation used in the main paper. This version is more simple to study, and we note that all nonlinear terms, except the SPM term, have some variant of $e^{i\omega_1\tau}$ multiplied onto them. This gives temporal oscillations on the carrier time scale. This is a consequence of the fact that even if this is an envelope approach, then it is actually the carrier that is modelled. We therefore stress that, e.g., the $A^*Ae^{i\omega_1 \tau}$ term should not be confused with optical rectification exactly because the $e^{i\omega_1 \tau}$ term is being retained.

The operator
\begin{align}\label{eq:Dhat}
\hat D_{\omega_1}=\sum_{m=2}m!^{-1}k_m(\omega_1)(i\partial_\tau)^m \end{align}
accounts for dispersion in time domain, which we conveniently evaluate directly in frequency domain as
\begin{align}\label{eq:Domega}
\tilde D_{\omega_1}(\omega)&=\sum_{m=2}m!^{-1}(\omega-\omega_1)^mk_m(\omega_1)
\\&
= k(\omega)-k_1(\omega_1)(\omega-\omega_1)-k(\omega_1)
\end{align}
without the need of a polynomial expansion.
The wavenumber $k(\omega)=n(\omega)\omega/c$, where $n(\omega)$ is the linear refractive index modelled by the $e$ polarized Sellmeier equation of 5\% MgO:LN \cite{gayer:2008}, and $k_m(\omega)=d^mk(\omega)/d\omega^m$ are the higher-order dispersion coefficients.
The $e^{\pm i\omega_1\tau}$ term on the nonlinear terms accounts for carrier-wave oscillations, and the peculiar term $\Delta_{\rm pg}=\omega_1 k_1(\omega_1)-k(\omega_1)=\omega_1(1/v_g-1/v_p)$ accounts for the phase-group-velocity mismatch (carrier-envelope phase slip), where $v_g=1/k_1(\omega_1)$ is the pump group velocity and $v_p=c/n(\omega_1)$ is the pump phase velocity. Finally, the $+$ sign implies that only the positive frequency content of the nonlinear term is used \cite{conforti:2010PRA}; we remind that the analytical field $A$ is defined over the entire frequency range $\omega\in [-\infty,\infty]$.

\section{Phase-matching conditions for the resonant radiation waves}

The theory for the RR phase matching conditions also requires using the NAEE model. 
%
The equations will support a number of phase-matching conditions between a soliton at the pump frequency and a linear (dispersive) wave.
The soliton envelope is the exact nonlinear solution in presence of SPM and GVD only. Particular to the case we study here, the soliton exists due to a self-defocusing \textit{effective} nonlinearity, which given by the sum of the Kerr SPM nonlinearity and the cascading nonlinearity. Let us for simplicity denote it $\kappa_{\rm eff}^{(3)}= \kappa_{\rm casc}^{(3)}+\kappa^{(3)}$, and the self-defocusing nature of the nonlinearity implies that $\kappa_{\rm eff}^{(3)}<0$. Thus, the ansatz $A(\zeta,\tau)=F_s(\tau)e^{i q_s\zeta}$, where $F_s$ is the soliton envelope (which is real), $q_s$ is the nonlinear wavenumber of the soliton, solves 
the following self-defocusing nonlinear Schr{\"o}dinger equation (see also \cite{bache:2010e} for more details)
\begin{align}\label{eq:NLSE}
i\partial_\zeta A -\tfrac{1}{2}k_2(\omega_s)(\partial_{\tau})^2A+
\kappa_{\rm eff}^{(3)}|A|^2A=0
\end{align}
A consequence of the self-defocusing negative nonlinearity is that $q_s<0$ \cite{bache:2010e}. Another direct consequence is the requirement of normal dispersion, $k_2(\omega_s)>0$.

To find the RR phase-matching conditions, we take the extended ansatz \cite{skryabin:2005} $A(\zeta,\tau)=F_s(\tau)e^{i q_s\zeta}+g(\zeta,\tau)$, where $g$ is the dispersive wave. To leading order we get
\begin{multline}\label{eq:PM-NAEE}
(i\partial_\zeta +\hat D_{\omega_s}) g
\\+\kappa^{(2)}F_s
[
 g e^{i\zeta(q_s-\Delta _{\rm pg})-i\omega_s \tau}
+2{\rm Re}(g e^{-iq_s\zeta}) e^{i\omega_s \tau+i\Delta _{\rm pg}\zeta}
]
\\
+
\kappa^{(3)}F_s^2
[g^*e^{i2q_s\zeta}+(2g^*+g)e^{i\zeta(-q_s+ 2\Delta _{\rm pg})+i2\omega_s\tau }
\\
+ge^{i\zeta(2q_s- 2\Delta _{\rm pg}) -i2\omega_s\tau }
]
=\\
-
\sum_{m=3}m!^{-1}k_m(\omega_s)(i\partial_\tau)^m
F_se^{iq_s\zeta} +\kappa^{(3)}_{\rm casc}F_s^3e^{iq_s\zeta}\\
-\kappa^{(2)}F_s^2 [e^{i\zeta(2q_s-\Delta _{\rm pg})-i\omega_s \tau}
+e^{i\omega_s \tau+i\Delta _{\rm pg}\zeta}
]
\\
-\kappa^{(3)}F_s^3[e^{i\zeta(-q_s+2\Delta _{\rm pg})+i2\omega_s \tau}
e^{i\zeta(3q_s-2\Delta _{\rm pg})-i2\omega_s \tau}]
\end{multline}

The next step is to find solutions for the dispersive wave $g$. We can make the ansatz $g(\zeta,\tau)=g'(\zeta,\tau) e^{i\tilde D_{\omega_s} \zeta-i(\omega-\omega_s)\tau}$, and after neglecting the nonlinear contributions proportional to $F_s$ and $F_s^2$ on the left-hand side, we get the phase matching conditions relating the dispersion on the left-hand side with the nonlinear driving terms on the right-hand side.

For the 4WM terms mediated by the $\chi^{(3)}$ nonlinear terms, the phase-matching conditions are well-known \cite{Conforti2013}
\begin{align}
\tilde D_{\omega_s}(\omega)&=q_s,  &(\text{SPM-RR}, \; |A|^2A)\\
\tilde D_{\omega_s}(\omega)&=-q_s+2\Delta_{\rm pg},  &(\text{SPM-cRR}, \; {|A|}^2A^*)\\
\tilde D_{\omega_s}(\omega)&=3q_s-2\Delta_{\rm pg},  &(\text{THG-RR}, \; {A^3})
\end{align}
The first condition is the traditional RR induced by SPM, the second is the "conjugate RR" from SPM (a.k.a. negative-frequency RR), while the third is the "third-harmonic generation RR" or simply THG-RR. We note that in quadratic nonlinear crystals, the SPM-RR has been predicted \cite{bache:2010e} and experimentally observed \cite{Zhou:2015,Zhou:2015-OL}, and the SPM-cRR has been studied numerically \cite{conforti-PhysRevA.88.013829-2013}.

Usually the soliton ansatz will remove the term $\propto F_s^3$ in Eq. (\ref{eq:PM-NAEE}), see e.g. \cite[Eq. (9)]{Conforti2013}. In our case it remains with the prefactor $\kappa_{\rm casc}^{(3)}$ because we consider the soliton ansatz that solves a self-defocusing NLSE with a reduced effective nonlinearity, Eq. (\ref{eq:NLSE}). However, it does not change anything for the 4WM phase-matching conditions since its phase-matching condition is identical to the SPM-RR case.

In the same way, our analysis here shows that the 3WM from the $\chi^{(2)}$ nonlinear terms will support the following phase-matching conditions
\begin{align}
\label{eq:PM-SFG-moving-suppl}
\tilde D_{\omega_s}(\omega)&=2q_s-\Delta_{\rm pg},  &(\text{SFG-RR}, \; {A^2})\\
\label{eq:PM-DFG-moving-suppl}
\tilde D_{\omega_s}(\omega)&=\Delta_{\rm pg},  &(\text{DFG-RR}, \; {A^*A})
\end{align}
We note here that there are no "conjugate" RR terms for the 3WM case: the term $A^*A$ is its own conjugate, and the conjugate of $A^2$ resides for negative frequencies only \cite{conforti:2010PRA} and is therefore not giving any relevant phase-matching conditions for $\omega>0$ [and also this is why it does not appear in Eq. (\ref{eq:NAEE-suppl})].

Note that the nonlinear wavenumber of the soliton $q_s$ is not entering the DFG phase-matching condition Eq. (\ref{eq:PM-DFG-moving-suppl}) because as we show below it cancels out as a result of the DFG mixing between the two soliton photons. We stress that it is not an indication that a soliton is not part of the phase-matching condition. In fact a similar effect is well known from 4WM RR, where in the nondegenerate case of a soliton and a linear probe interacting, $q_s$ may also cancel \cite[$J=+1$ case in Eq. (11)]{skryabin:2005}.

In all the non-standard cases, the term $\Delta_{\rm pg}$ appears, and it is therefore important to specify that in this context we intend it to be evaluated at the soliton frequency $\omega_s$, i.e. specifically $\Delta_{\rm pg}=\omega_s/v_{g,s}-k(\omega_s)$, where $v_{g,s}=1/k_1(\omega_s)$ is the soliton group velocity.

For a physical interpretation it is instructive to transform the interaction back to the lab-frame coordinate, because this reveals the direct wave-number phase-matching conditions. In this connection, it is instructive to mention that the soliton dispersion relation is $k_s(\omega)=k(\omega_s)+(\omega-\omega_s)/v_{g,s}+q_s$.
For the 4WM we get
\begin{align}
\label{eq:PM-SPM}
  k(\omega)&= k_s(\omega),  &(\text{SPM-RR}, \; {|A|^2A})\\
\label{eq:PM-cSPM}
  k(\omega)&=-k_s(\omega)+2\omega/v_{g,s},  &(\text{cSPM-RR}, \; {|A|^2A^*}) \\
\label{eq:PM-TH}
  k(\omega)&=3k_s(\omega)-2\omega/v_{g,s},  &(\text{THG-RR}, \; {A^3})
\end{align}
For the 3WM we get
\begin{align}
\label{eq:PM-SFG}
  k(\omega)&=2k_{s}(\omega)-\omega/v_{g,s},  &(\text{SFG-RR}, \; {A^2})\\
\label{eq:PM-DFG}
  k(\omega)&=\omega/v_{g,s},  &(\text{DFG-RR}, \; {A^*A})
\end{align}

Let us discuss these results, because except for Eq. (\ref{eq:PM-SPM}) they have not been reported in this form before. Exactly Eq. (\ref{eq:PM-SPM}) is therefore a good place to start: it simply means that the wavenumbers of the soliton and the dispersive wave match at the RR frequency. However, as mentioned above, all the other phase-matching equations have the term $\Delta_{\rm pg}$ and this leads to the $\omega_{\rm RR}/v_{g,s}$ terms in the above representations. As we show below, this term represents the group-velocity mismatch (GVM) between the soliton and the harmonic wave of the nonlinear process.

In the two other 4WM cases it is more involved: for the cSPM-RR the phase-matching condition is equivalent to
\begin{align}\label{eq:PM-cSPM-kvecs}
k(\omega)=-k_s(-\omega)
\end{align}
i.e., the RR wave is phase-matched to the soliton evaluated at the negative frequency of the RR wave. The explanation behind this peculiar "negative frequency RR" is that because an equivalent to Eq. (\ref{eq:NAEE-suppl}) exists expressed by the complex-conjugate system, we can express the "conjugate" negaton, i.e., the forward-propagating negaton that solves the complex conjugate equation system, as $k_s^c(\omega)=-k_s(-\omega)=-k(\omega_s)+(\omega+\omega_s)/v_{g,s}-q_s$, and hence $k(\omega)=k_s^c(\omega)$ gives an equivalent but positive phase-matching frequency \cite{rubino:2012,conforti-PhysRevA.88.013829-2013}.

For the THG-RR case, we get
\begin{align}\label{eq:PM-THG-kvecs}
k(\omega)=k_s(\omega_a)+k_s(\omega_b)+k_s(\omega_c)
\end{align}
which is the 4WM equivalent of an SFG process and for energy conservation we require $\omega_a +\omega_b+\omega_c=\omega$. The challenge for the THG-RR case is substantial: it turns out to be phase-matched deep in the low-frequency part of the spectrum \cite{Conforti2013}, but nonetheless the analysis here shows that all the contributing soliton photons must have lower frequencies. This perhaps explains why it has yet to be observed even in simulations.

Next let us consider Eq. (\ref{eq:PM-SFG}) that is a result of the $A^2$ wave mixing term. It is straight-forward to show that it is equivalent to
\begin{align}\label{eq:PM-SFG-kvecs}
k(\omega)=k_s(\omega_a)+k_s(\omega_b), \quad \omega_a+\omega_b=\omega
\end{align}
i.e., the SFG between soliton photons at two different frequencies, constrained of course with energy conservation to give the new frequency.
If we expand the dispersion on the left-hand side around the second-harmonic (SH) frequency of the soliton frequency $\omega_2=2\omega_s$, we get
\begin{align}
\tilde D_{\omega_2}(\omega)
&-(\omega-\omega_2)\dshg+\Delta k_{\rm SHG}-2q_s=0
\label{eq:PM-SHG-expansion}
\end{align}
where $\dshg=k_1(\omega_s)-k_1(2\omega_s)$ is the GVM coefficient between the soliton and its SH. $\Delta k_{\rm SHG}=k(2\omega_s)-2k(\omega_s)$ is the SHG phase mismatch between the soliton and its SH.
In \cite{zhou:PhysRevA.90.013823} we used an alternative route to arrive at a similar result, exploiting the coupled-wave equations for the pump and its SH in the slow-varying envelope approximation. In the theory reported there, we for simplicity only considered up to 2. order dispersion, $m=2$ in $\tilde D_{\omega_2}(\omega)$ in Eq. (\ref{eq:PM-SHG-expansion}), and neglected the soliton nonlinear phase $q_s$. The phase-matching conditions were then found by setting the denominator in Eq. (8) in \cite{zhou:PhysRevA.90.013823} to zero, which is precisely the condition reported above in Eq. (\ref{eq:PM-SHG-expansion}). Thus, the phase matching conditions reported in Eq. (9) \cite{zhou:PhysRevA.90.013823} are exactly the SFG-RR phase-matching condition reported above in Eq. (\ref{eq:PM-SHG-expansion}). In \cite{zhou:PhysRevA.90.013823} we explain the resonances as a result of the "residual" SH dispersion operator experiencing phase-matching, which happens when operating in the regime strongly detuned from the SH phase-matching condition $\Delta k_{\rm SHG}=0$ and when a soliton is excited. This is connected to the so-called nonlocal resonances, first predicted in cascaded SHG by some of us \cite{bache:2007a}.

In the DFG-RR case we can rewrite the phase-matching condition Eq. (\ref{eq:PM-DFG}) as
\begin{align}\label{eq:PM-DFG-kvecs}
k(\omega)=k_s(\omega_a)-k_s(\omega_b), \quad \omega_a-\omega_b=\omega
\end{align}
i.e., the DFG between soliton photons at two different frequencies, constrained of course with energy conservation so the difference between them gives the new frequency. When expanding the linear wave dispersion on the right-hand side around some low-frequency value $\omega_{\rm DFG}$ the condition becomes
\begin{align}
\tilde D_{\omega_{\rm DFG}}(\omega)
&-(\omega-\omega_{\rm DFG})\ddfg-\Delta k_{\rm DFG}=0
\label{eq:PM-DFG-expansion}
\end{align}
where $\ddfg=k_1(\omega_s)-k_1(\omega_{\rm DFG})$ is the GVM coefficient between the soliton and the DFG frequency. The DFG phase-mismatch coefficient is given by
\begin{align}\label{eq:DeltaDFG}
\Delta k_{\rm DFG}=k_s(\omega_s)-k_s(\omega_s-\omega_{\rm DFG})-k(\omega_{\rm DFG})
\end{align}
i.e., the DFG wave is a result of 3WM through DFG between the soliton at $\omega_s$ and the soliton at a detuned frequency $\omega_s-\omega_{\rm DFG}$. Note that $\Delta k_{\rm DFG}\neq 0$ (in fact, in a type-0 interaction in a crystal like LN we will find that we always have $\Delta k_{\rm DFG}< 0$ just like we always have $\Delta k_{\rm SHG}> 0$), so the DFG process is heavily phase-mismatched. The RR wave will then appear at the frequency where Eq. (\ref{eq:PM-DFG-expansion}) is zero.

The conclusion of this is that we can understand the 3WM phase-matching conditions of the RR waves as a consequence of the soliton not finding direct phase-matching because $\Delta k\neq 0$, and instead some new frequency becomes phase-matched due to GVM and higher-order dispersion effects. Additionally, the phase-matching condition is directly linked to the so-called nonlocal response of the cascaded 3WM \cite{bache:2007a}, where the RR phase-matching condition is equivalent to the case where nonlocal response function $\tilde R(\Omega)$, see e.g. \cite[Eq. (6)]{bache:2008}, will have poles in the denominator.

Getting back to the form of the DFG-RR phase-matching condition reported in Eq. (\ref{eq:PM-DFG}), we can write the left-hand side as $\omega/v_{\rm ph}(\omega)$, where $v_{\rm ph}(\omega)=c/n(\omega)$ is the phase velocity at the frequency $\omega$. In this way, the DFG-RR condition becomes a very particular requirement, namely that
\begin{align}\label{eq:PM-DFG-v-suppl}
v_{\rm ph}(\omega)=v_{g,s}
\end{align}
i.e., that the phase velocity of the RR wave is the same as the group velocity of the soliton. Alternatively it is expressed as $n(\omega)=n_{g,s}$ where $n_{g,s}=c/v_{g,s}$ is the group index of the soliton. Such a condition is well-known from THz generation through optical rectification \cite{Nahata1996}, where it is known as the \textit{velocity-matching condition}. It is not easy to fulfill this condition because the phase- and group-velocities are quite different even when considering that the RR wave is allowed to have any frequency within the transparency region of the crystal. For a fixed soliton frequency it is possible to achieve velocity-matching, i.e. fulfill condition Eq. (\ref{eq:PM-DFG-v-suppl}), when the RR frequency is higher than the soliton frequency. This because within a certain transparency region all materials have for a fixed frequency the phase index $n$ below the group-index $n_g$, i.e. that the phase velocity is faster than the group velocity, and additionally both will monotonically increase with frequency. Consequently, the soliton must necessarily look towards higher frequencies to find a wave with an phase-index of the same value. However, through the analysis presented above it is a requirement that the RR frequency is located to the red side of the soliton, $\omega_{\rm RR}<\omega_s$, otherwise the soliton wavenumbers do not cancel. This requirement practically makes velocity-matching impossible in LN, unless one goes beyond an IR resonance and exploit that on the other side of the resonance in the far-IR transparency window the phase index is sufficiently high to achieve velocity matching. This is essentially what is done in the THz case.

\section{Quasi-phase matching control of the resonant radiation phase-matching conditions}

We here exploit the quasi-phase matching (QPM) technique to achieve velocity matching in the same transparency window as the soliton. QPM employs a periodic-poling structure of the quadratic nonlinearity, so we essentially impose a grating structure on the effective nonlinearity that is generally expressed as $d_{\rm eff} g_{\rm QPM}(z)$, where $g_{\rm QPM}(z)$ is the normalized QPM grating function. The simplest and most widely used case is where the grating is a square function that effectively reverses the sign of $\chi^{(2)}$ with 50\% duty cycle and periodicity $\Lambda$. Expressing the square grating in a Fourier series gives
\begin{align}
g_{\rm QPM}(z)&=\frac{4}{\pi}\sum_{l=0}^\infty \frac{1}{2l+1}\sin\left[
(2l+1)\kqpm z
\right]
\nonumber\\
\label{eq:QPM}
&=\frac{2}{\pi}\sum_{l=-\infty}^\infty \frac{-i}{2l+1}
e^{i(2l+1)\kqpm z}
\end{align}
where $\kqpm=2\pi/\Lambda$. We here immediately see the well-known $2/\pi$ prefactor, which is the "penalty" on the nonlinear strength for using a uniform QPM square-grating poling compared to the unpoled case. The general idea behind QPM is that the exponential terms $m_0^{-1}e^{im_0\kqpm z}$, $m_0=(2l+1)=\pm 1,\pm 3,\pm 5,\ldots$, contribute to the similar exponential terms in front of the $\kappa^{(2)}$ terms, here $e^{\pm i \Delta_{\rm pg}\zeta}$, respectively. In principle there is an infinite series of contributions when written in terms of the exponential expansion. However, we also see that the $m_0^{-1}$ coefficient makes the higher-order terms irrelevant as the nonlinear strength quickly drops for increasing $m_0$ values. Therefore it is custom to consider only the first few orders to see if phase-matching can be achieved.

Therefore, using a QPM square grating the 3WM phase-matching conditions change to
\begin{align}
\label{eq:PM-SFG-moving-QPM-suppl}
\tilde D_{\omega_s}(\omega)&=2q_s-\Delta_{\rm pg}+ \kqpm, \; &(\text{SFG-RR}, \; {A^2})\\
\label{eq:PM-DFG-moving-QPM-suppl}
\tilde D_{\omega_s}(\omega)&=\Delta_{\rm pg}-\kqpm, \; &(\text{DFG-RR}, \; {A^*A})
\end{align}
In principle there is an $m_0$ term in front of the $\kqpm$. However, we have here used the knowledge that $\Delta_{\rm pg}>0$ so that the SFG case needs QPM to increase the right-hand side of (\ref{eq:PM-SFG-moving-QPM-suppl}), thus invoking the $+\kqpm$ term of the exponential QPM grating expansion, and similarly for the DFG case we choose the $-\kqpm$ term. Expressed in the stationary lab frame we get
\begin{align}
\label{eq:PM-SFG-QPM}
  k(\omega)&=2k_{s}(\omega)+\kqpm-\omega/v_{g,s},  &(\text{SFG-RR}, \; {A^2})\\
\label{eq:PM-DFG-QPM}
  k(\omega)&=-\kqpm+\omega/v_{g,s},  &(\text{DFG-RR}, \; {A^*A})
\end{align}
At this stage, it is a matter of finding the right grating pitch $\Lambda$ to achieve RR wave phase matching.

We should also emphasize that RR wave generation through 3WM is quite powerful because it gives a parametrically tunable RR phase-matching frequency. In the SFG-RR case, one can tune the RR frequency through $k(\omega)$; as we have seen above this is essentially the SH wavenumber, and in a birefringent (type I) configuration this gives the opportunity to widely tune the RR frequency as shown in our recent experiment \cite{zhou:PhysRevA.90.013823}. In the type 0 case we investigate here, QPM is needed to do this, but it still gives a very powerful access to controlling the RR frequency, both in the SFG case and the DFG case. Such a parametrically tunable RR cannot be found in 4WM. The SPM cases simply do not offer this kind of control. While the THG case does in principle offer a similar kind of birefringent control of the TH wavenumber, i.e. the left-hand side of Eq. (\ref{eq:PM-TH}), if the soliton forms in a birefringent medium, the THG-RR case is very elusive and almost all relevant cases studied so far for 4WM are fibers or waveguides in nonlinear media that are not birefringent.

Using QPM to achieve velocity matching for THz wave generation has been implemented in lithium niobate \cite{Lee2000} (see review in \cite{Lhuillier2007a,Lhuillier2007}), but it has to our knowledge not been used to generate velocity matching in the same transparency window as the pump/soliton wave, i.e., in the primary VIS-IR transparency range of $0.3-5.5\mic$. This would also require a very broadband pump; in the standard case an 800 nm pump is used, and if we want to generate an RR at $5\mic$ then the pump bandwidth should be around 150 nm, corresponding to a sub-10 fs pump pulse.

\section{Numerical simulations}

The numerical simulations of Eq. (\ref{eq:NAEE-suppl}) were performed using a plane-wave split-step Fourier method, written in the interaction picture and employing a variable step size ODE solver (ode23 in Matlab). We found a relative tolerance of $10^{-6}$ crucial to achieve stability of the high harmonics. The actual equation that we solved was rewritten somewhat, details can be found in \cite[Eq. (31)]{bache:2016-NAEE}.

The challenge of a code resolving the carrier wave is to get the temporal resolution fine enough so that all the (relevant) interacting frequencies have enough temporal resolution. We found that in the system a QPM resonance gave significant radiation around the third harmonic (500-600 nm range), and if the carrier wave of this frequency has to be temporally resolved we should use well below 1 fs time resolution. We typically used a 100 as time grid (corresponding to simulating up to the $25^{\rm th}$ harmonic of the pump) and $2^{16}$ or $2^{17}$ grid size. The latter was largely determined by the group-velocity mismatch between the pump wave and the generated supercontinuum, i.e. whether the generated waves remain in the spectral window. Conversely this in spectral domain leads to a requirement of a high spectral resolution, i.e. exactly a requirement of a large number of grid points for a fixed bandwidth.

The simulations used as an initial condition quantum noise seeds corresponding to the Wigner representation (on average 1/2 photon per discrete time grid, see \cite{Zhou2016-aplp} for details). Randomizing this noise seed gave us the opportunity to calculate the complex first order degree of coherence.

Fig. 1 in the main paper shows an excellent quantitative agreement with the simulation and the experiment. To achieve this agreement we fixed the Kerr and Raman nonlinear parameters, as found in our recent work \cite{Zhou:2015}: this implies that $f_R=0.35$ was used and the Kerr nonlinearity was only modified slightly using Miller's scaling to account for the slightly longer pump wavelength used here, giving $n_2=52\cdot 10^{-20}~\rm m^2/W$ at $\lambda_0=1.75\mic$. We then adjusted the quadratic nonlinear tensor strength $d_{33}$ within a reasonable parameter space to look for agreement with the experiment, and found that $d_{33}=18.0$ pm/V gave an extremely convincing agreement, not just for the presented plot but for the entire parameter range explored experimentally. To justify this choice, Shoji et al. \cite{Shoji:1997} originally found for 5\%MgO doped congruent LN (1) $d_{33}=25.0$ pm/V at 1064 nm and (2) $d_{33}=20.3$ pm/V at 1310 nm (both with 10\% uncertainty). Such a big difference cannot be explained by Miller's scaling, and it is not clear why such different values were obtained and which one is more accurate. If we use Miller's scaling to go to the pump wavelength range we used here ($1.6$-$1.8\mic$), we should get (1) $d_{33}=22$ pm/V or (2) $d_{33}=19$ pm/V, both with 10\% uncertainty. The somewhat reduced value we used in the simulations can then easily be justified as a result of imperfect poling or simply just experimental error on the determination of the nonlinearity. Gayer et al. \cite{gayer:2008} found that PPLN samples from various manufacturers had quite different effective QPM nonlinearities (i.e. $\tfrac{2}{\pi}d_{33}$ values), indicating that the poling quality can vary quite a lot. In the experiment we used 3 different PPLN manufacturers, namely Covesion, HC Photonics, and custom made samples. The data shown in the Fig. 1, 2 and 3(a) in the paper were from a Covesion PPLN crystal, while the 20 mm crystal used for Fig. 3(b) was from HC Photonics. The custom-made samples were used for initial experiments to confirm the concept.

\begin{figure}[tb]
\centerline{\includegraphics[width=\linewidth]{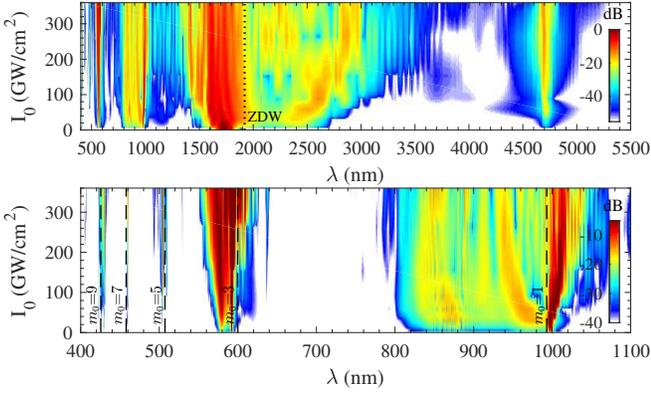}}
\caption{Numerical simulations corresponding to the same data of Fig. 2 in the main paper. The normalized spectra are shown in a false 2D color plot during an intensity sweep from 0-360 ${\rm GW/cm^2}$ with exactly the same parameters as used in the experiment. The bottom plot focuses on details in the visible and short-wavelength near-IR range, including calculated QPM resonances (dashed lines).
}
\label{fig:sim}
\end{figure}

Fig. \ref{fig:sim} shows the results of numerical simulations for an intensity sweep with exactly the same parameters as in Fig. 2 in the main paper. The main differences seem to be that the DFG-RR wave is not as broadband as in the experiment. The SPM-RR plateau is also more structured, but the overall trend is extremely similar to the experiment. The near-IR SFG-RR has interference fringes for high intensities, which we did not see experimentally, although we have to add that the spectral resolution in the experiment in this range is 7 nm so some of these fine features were hard to measure. More striking is probably that just on the blue side of the SFG-RR wave another wave emerges that becomes blue-shifted with increasing intensity. This was not observed experimentally. We also found that in the simulations the TH and the QPM resonances were significantly stronger than in the experiment. Note that simulations using the coupled slowly-varying envelope equations would not show any DFG-RR wave as the DFG term is discarded.

\begin{figure}[tb]
\centerline{\includegraphics[width=\linewidth]{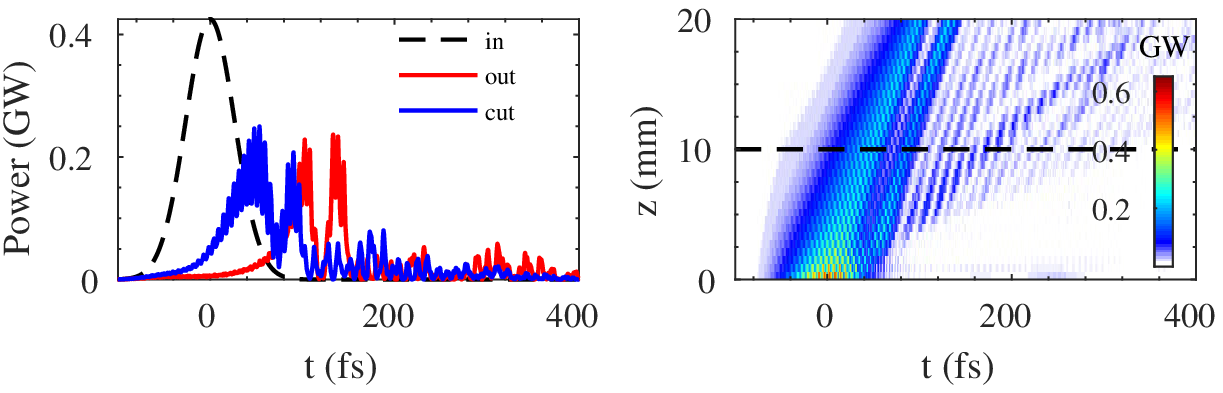}}
\centerline{\includegraphics[width=\linewidth]{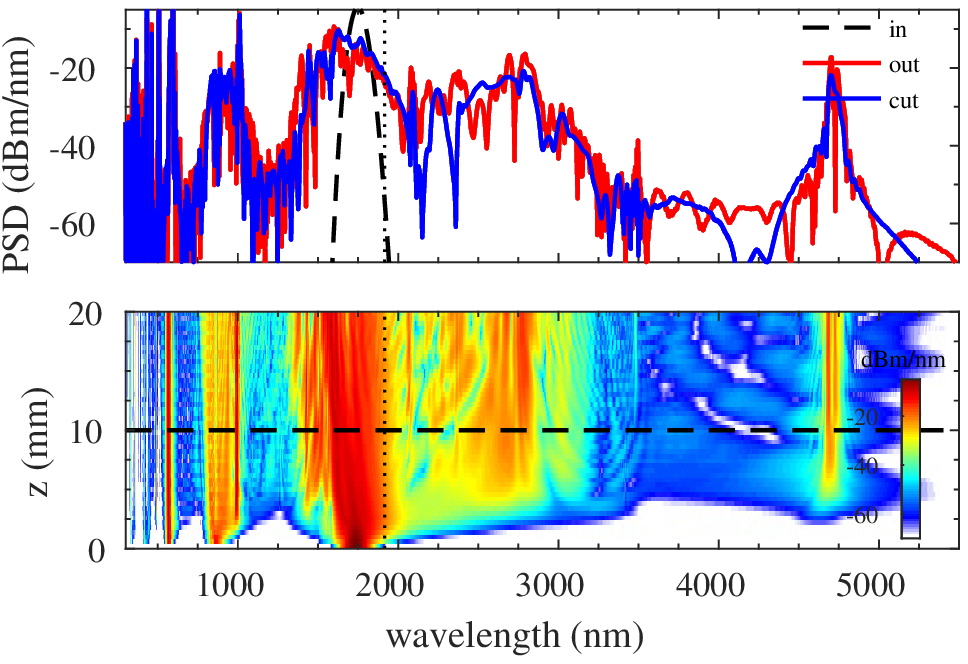}}
\caption{Numerical simulations behind the case shown in Fig. 1 in the main part. Here the evolution along the crystal is shown up to 20 mm PPLN with $\Lambda=30.0\mic$. Parameters: $\lambda_0=1.75\mic$, 150 $\rm GW/cm^2$, 60 fs FWHM Gaussian pulse duration and 80 nm FWHM bandwidth (input pulse slightly chirped, which was taken linear and negative, GDD$=-400~\rm fs^2$), $n_{2,\rm Kerr}=52\cdot 10^{-20}~\rm m^2/W$, $d_{\rm eff}=18$ pm/V. $2^{17}$ grid points were used with a frequency window spanning up to the $25^{\rm th}$ harmonic. The PSD is calculated based on a 1 kHz repetition rate and a Gaussian spot size of 0.5 mm FWHM.
}
\label{fig:sim_intensity_150}
\end{figure}

In Fig. \ref{fig:sim_intensity_150} the full simulation behind Fig. 1 in the main paper is shown, and the pulse is propagated up to 20 mm length. The soliton forms on the leading edge of the pulse, while the trailing edge has a strongly asymmetric form. This is due to self-steepening of the leading edge, and eventually a soliton also forms there. This gives a pulse splitting effect, which is most likely due to a combination of the competing Raman nonlinearity and strong self-steepening.

\begin{figure}[tb]
\centerline{\includegraphics[width=\linewidth]{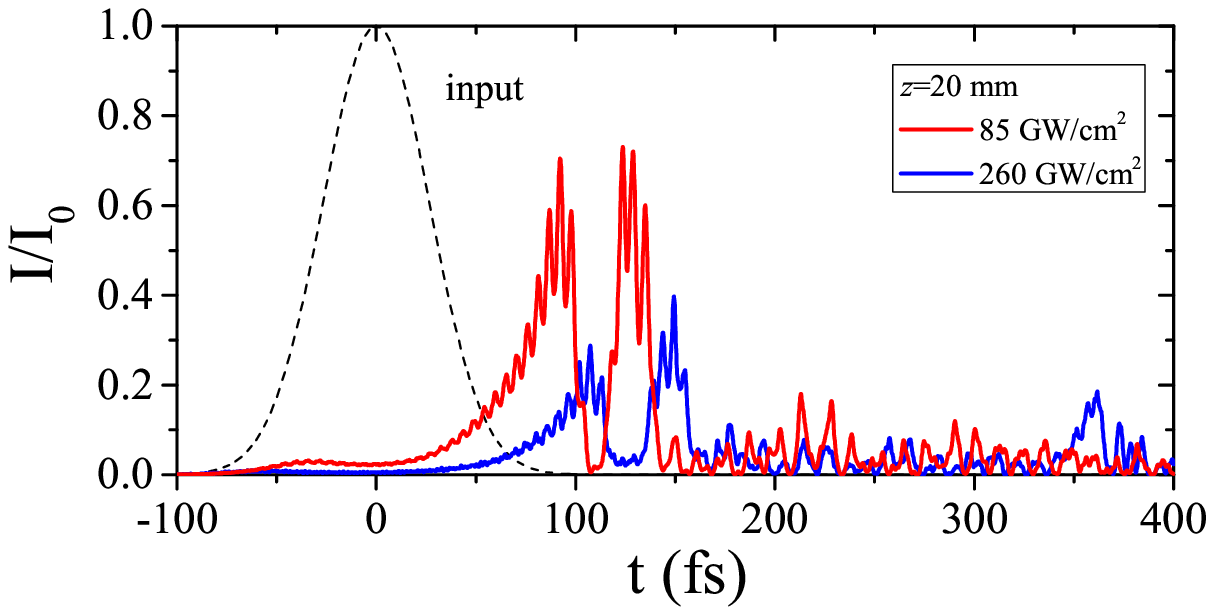}}
\centerline{\includegraphics[width=\linewidth]{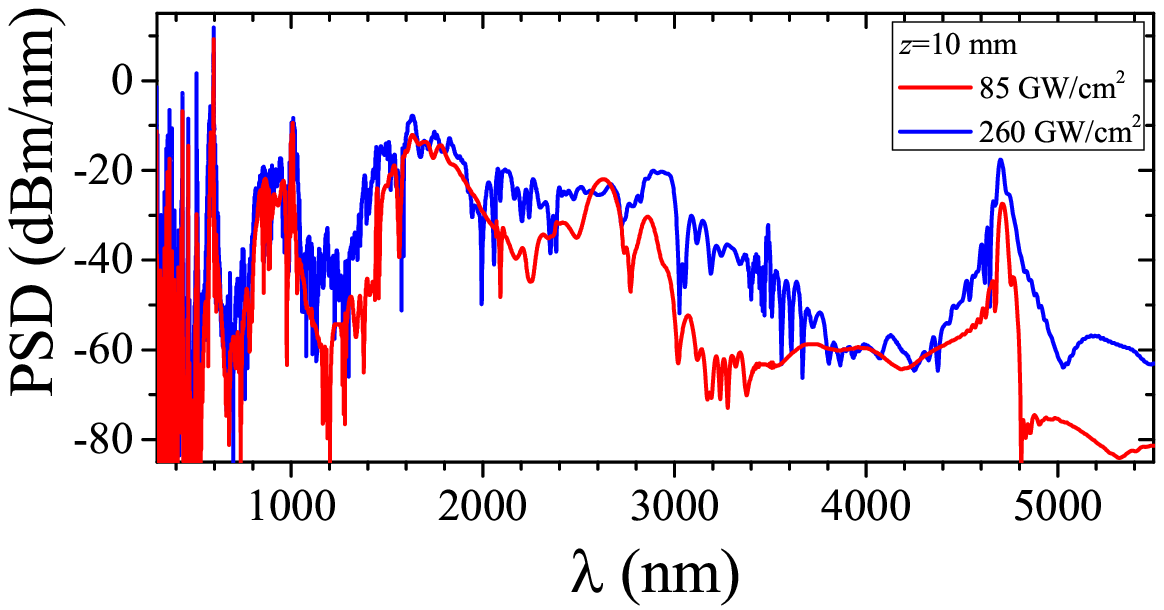}}
\caption{Numerical simulations for the same basic parameters as Fig. \ref{fig:sim_intensity_150}, but where a medium intensity case ($I_0=85~\rm GW/cm^2$) and high intensity case ($I_0=260 ~\rm GW/cm^2$) are compared in time (at $z=20$ mm) and spectrum (at $z=10$ mm).
}
\label{fig:sim_intensity_85_260}
\end{figure}

In the experiment we observed that the SPM-RR plateau flattened for high intensities. Fig. \ref{fig:sim_intensity_85_260} gives some insight into this by comparing a medium-intensity simulation, 85 $\rm GW/cm^2$, with a high-intensity simulation, 260 $\rm GW/cm^2$. It seems that for high intensities the many 3WM and 4WM processes give rise to substantial frequency conversion, especially to high frequencies. This depletes the pump and thus weakens the soliton. This is clearly seen in the time plot, where the soliton at high intensity is much weaker than the input pulse when comparing with the medium intensity case. On top of that, it is well known that close to phase matching the cascading has a significant self-steepening contribution, which increases significantly for high intensities. This again leads to a different dynamics concerning the formation of the SPM-RR wave. We see in the comparison of the spectra at 10 mm that the high-intensity case has a much  flatter plateau in the region from 2.0-$3.5\mic$, where the SPM-RR is formed. It also extends further into the mid-IR. This agrees well with the experimental data.



\begin{thebibliography}{57}%
\makeatletter
\providecommand \@ifxundefined [1]{%
 \@ifx{#1\undefined}
}%
\providecommand \@ifnum [1]{%
 \ifnum #1\expandafter \@firstoftwo
 \else \expandafter \@secondoftwo
 \fi
}%
\providecommand \@ifx [1]{%
 \ifx #1\expandafter \@firstoftwo
 \else \expandafter \@secondoftwo
 \fi
}%
\providecommand \natexlab [1]{#1}%
\providecommand \enquote  [1]{``#1''}%
\providecommand \bibnamefont  [1]{#1}%
\providecommand \bibfnamefont [1]{#1}%
\providecommand \citenamefont [1]{#1}%
\providecommand \href@noop [0]{\@secondoftwo}%
\providecommand \href [0]{\begingroup \@sanitize@url \@href}%
\providecommand \@href[1]{\@@startlink{#1}\@@href}%
\providecommand \@@href[1]{\endgroup#1\@@endlink}%
\providecommand \@sanitize@url [0]{\catcode `\\12\catcode `\$12\catcode
  `\&12\catcode `\#12\catcode `\^12\catcode `\_12\catcode `\%12\relax}%
\providecommand \@@startlink[1]{}%
\providecommand \@@endlink[0]{}%
\providecommand \url  [0]{\begingroup\@sanitize@url \@url }%
\providecommand \@url [1]{\endgroup\@href {#1}{\urlprefix }}%
\providecommand \urlprefix  [0]{URL }%
\providecommand \Eprint [0]{\href }%
\providecommand \doibase [0]{http://dx.doi.org/}%
\providecommand \selectlanguage [0]{\@gobble}%
\providecommand \bibinfo  [0]{\@secondoftwo}%
\providecommand \bibfield  [0]{\@secondoftwo}%
\providecommand \translation [1]{[#1]}%
\providecommand \BibitemOpen [0]{}%
\providecommand \bibitemStop [0]{}%
\providecommand \bibitemNoStop [0]{.\EOS\space}%
\providecommand \EOS [0]{\spacefactor3000\relax}%
\providecommand \BibitemShut  [1]{\csname bibitem#1\endcsname}%
\let\auto@bib@innerbib\@empty
\bibitem [{\citenamefont {Mollenauer}\ \emph {et~al.}(1980)\citenamefont
  {Mollenauer}, \citenamefont {Stolen},\ and\ \citenamefont
  {Gordon}}]{mollenauer:1980}%
  \BibitemOpen
  \bibfield  {author} {\bibinfo {author} {\bibfnamefont {L.~F.}\ \bibnamefont
  {Mollenauer}}, \bibinfo {author} {\bibfnamefont {R.~H.}\ \bibnamefont
  {Stolen}}, \ and\ \bibinfo {author} {\bibfnamefont {J.~P.}\ \bibnamefont
  {Gordon}},\ }\bibfield  {title} {\enquote {\bibinfo {title} {Experimental
  observation of picosecond pulse narrowing and solitons in optical fibers},}\
  }\href {\doibase 10.1103/PhysRevLett.45.1095} {\bibfield  {journal} {\bibinfo
   {journal} {Phys. Rev. Lett.}\ }\textbf {\bibinfo {volume} {45}},\ \bibinfo
  {pages} {1095--1098} (\bibinfo {year} {1980})}\BibitemShut {NoStop}%
\bibitem [{\citenamefont {Wai}\ \emph {et~al.}(1986)\citenamefont {Wai},
  \citenamefont {Menyuk}, \citenamefont {Lee},\ and\ \citenamefont
  {Chen}}]{Wai:1986}%
  \BibitemOpen
  \bibfield  {author} {\bibinfo {author} {\bibfnamefont {P.~K.~A.}\
  \bibnamefont {Wai}}, \bibinfo {author} {\bibfnamefont {C.~R.}\ \bibnamefont
  {Menyuk}}, \bibinfo {author} {\bibfnamefont {Y.~C.}\ \bibnamefont {Lee}}, \
  and\ \bibinfo {author} {\bibfnamefont {H.~H.}\ \bibnamefont {Chen}},\
  }\bibfield  {title} {\enquote {\bibinfo {title} {Nonlinear pulse propagation
  in the neighborhood of the zero-dispersion wavelength of monomode optical
  fibers},}\ }\href {\doibase 10.1364/OL.11.000464} {\bibfield  {journal}
  {\bibinfo  {journal} {Opt. Lett.}\ }\textbf {\bibinfo {volume} {11}},\
  \bibinfo {pages} {464--466} (\bibinfo {year} {1986})}\BibitemShut {NoStop}%
\bibitem [{\citenamefont {Akhmediev}\ and\ \citenamefont
  {Karlsson}(1995)}]{akhmediev:1995}%
  \BibitemOpen
  \bibfield  {author} {\bibinfo {author} {\bibfnamefont {Nail}\ \bibnamefont
  {Akhmediev}}\ and\ \bibinfo {author} {\bibfnamefont {Magnus}\ \bibnamefont
  {Karlsson}},\ }\bibfield  {title} {\enquote {\bibinfo {title} {Cherenkov
  radiation emitted by solitons in optical fibers},}\ }\href {\doibase
  10.1103/PhysRevA.51.2602} {\bibfield  {journal} {\bibinfo  {journal} {Phys.
  Rev. A}\ }\textbf {\bibinfo {volume} {51}},\ \bibinfo {pages} {2602--2607}
  (\bibinfo {year} {1995})}\BibitemShut {NoStop}%
\bibitem [{\citenamefont {Skryabin}\ and\ \citenamefont
  {Gorbach}(2010)}]{Skryabin:2010}%
  \BibitemOpen
  \bibfield  {author} {\bibinfo {author} {\bibfnamefont {Dmitry~V.}\
  \bibnamefont {Skryabin}}\ and\ \bibinfo {author} {\bibfnamefont {Andrey~V.}\
  \bibnamefont {Gorbach}},\ }\bibfield  {title} {\enquote {\bibinfo {title}
  {Colloquium: Looking at a soliton through the prism of optical
  supercontinuum},}\ }\href {\doibase 10.1103/RevModPhys.82.1287} {\bibfield
  {journal} {\bibinfo  {journal} {Rev. Mod. Phys.}\ }\textbf {\bibinfo {volume}
  {82}},\ \bibinfo {pages} {1287--1299} (\bibinfo {year} {2010})}\BibitemShut
  {NoStop}%
\bibitem [{\citenamefont {Dudley}\ \emph {et~al.}(2006)\citenamefont {Dudley},
  \citenamefont {Genty},\ and\ \citenamefont {Coen}}]{dudley:2006}%
  \BibitemOpen
  \bibfield  {author} {\bibinfo {author} {\bibfnamefont {John~M.}\ \bibnamefont
  {Dudley}}, \bibinfo {author} {\bibfnamefont {Go\"ery}\ \bibnamefont {Genty}},
  \ and\ \bibinfo {author} {\bibfnamefont {St\'ephane}\ \bibnamefont {Coen}},\
  }\bibfield  {title} {\enquote {\bibinfo {title} {Supercontinuum generation in
  photonic crystal fiber},}\ }\href {\doibase 10.1103/RevModPhys.78.1135}
  {\bibfield  {journal} {\bibinfo  {journal} {Rev. Mod. Phys.}\ }\textbf
  {\bibinfo {volume} {78}},\ \bibinfo {eid} {1135} (\bibinfo {year}
  {2006})}\BibitemShut {NoStop}%
\bibitem [{\citenamefont {Joly}\ \emph {et~al.}(2011)\citenamefont {Joly},
  \citenamefont {Nold}, \citenamefont {Chang}, \citenamefont {H\"olzer},
  \citenamefont {Nazarkin}, \citenamefont {Wong}, \citenamefont {Biancalana},\
  and\ \citenamefont {Russell}}]{joly:2011}%
  \BibitemOpen
  \bibfield  {author} {\bibinfo {author} {\bibfnamefont {N.~Y.}\ \bibnamefont
  {Joly}}, \bibinfo {author} {\bibfnamefont {J.}~\bibnamefont {Nold}}, \bibinfo
  {author} {\bibfnamefont {W.}~\bibnamefont {Chang}}, \bibinfo {author}
  {\bibfnamefont {P.}~\bibnamefont {H\"olzer}}, \bibinfo {author}
  {\bibfnamefont {A.}~\bibnamefont {Nazarkin}}, \bibinfo {author}
  {\bibfnamefont {G.~K.~L.}\ \bibnamefont {Wong}}, \bibinfo {author}
  {\bibfnamefont {F.}~\bibnamefont {Biancalana}}, \ and\ \bibinfo {author}
  {\bibfnamefont {P.~St.~J.}\ \bibnamefont {Russell}},\ }\bibfield  {title}
  {\enquote {\bibinfo {title} {Bright spatially coherent wavelength-tunable
  deep-{UV} laser source using an {Ar}-filled photonic crystal fiber},}\ }\href
  {\doibase 10.1103/PhysRevLett.106.203901} {\bibfield  {journal} {\bibinfo
  {journal} {Phys. Rev. Lett.}\ }\textbf {\bibinfo {volume} {106}},\ \bibinfo
  {pages} {203901} (\bibinfo {year} {2011})}\BibitemShut {NoStop}%
\bibitem [{\citenamefont {Mak}\ \emph {et~al.}(2013)\citenamefont {Mak},
  \citenamefont {Travers}, \citenamefont {H\"{o}lzer}, \citenamefont {Joly},\
  and\ \citenamefont {Russell}}]{Mak2013}%
  \BibitemOpen
  \bibfield  {author} {\bibinfo {author} {\bibfnamefont {Ka~Fai}\ \bibnamefont
  {Mak}}, \bibinfo {author} {\bibfnamefont {John~C.}\ \bibnamefont {Travers}},
  \bibinfo {author} {\bibfnamefont {Philipp}\ \bibnamefont {H\"{o}lzer}},
  \bibinfo {author} {\bibfnamefont {Nicolas~Y.}\ \bibnamefont {Joly}}, \ and\
  \bibinfo {author} {\bibfnamefont {Philip St.~J.}\ \bibnamefont {Russell}},\
  }\bibfield  {title} {\enquote {\bibinfo {title} {Tunable vacuum-{UV} to
  visible ultrafast pulse source based on gas-filled kagome-{PCF}},}\ }\href
  {\doibase 10.1364/OE.21.010942} {\bibfield  {journal} {\bibinfo  {journal}
  {Opt. Express}\ }\textbf {\bibinfo {volume} {21}},\ \bibinfo {pages}
  {10942--10953} (\bibinfo {year} {2013})}\BibitemShut {NoStop}%
\bibitem [{\citenamefont {Belli}\ \emph {et~al.}(2015)\citenamefont {Belli},
  \citenamefont {Abdolvand}, \citenamefont {Chang}, \citenamefont {Travers},\
  and\ \citenamefont {Russell}}]{Belli:2015-VUV-DW}%
  \BibitemOpen
  \bibfield  {author} {\bibinfo {author} {\bibfnamefont {Federico}\
  \bibnamefont {Belli}}, \bibinfo {author} {\bibfnamefont {Amir}\ \bibnamefont
  {Abdolvand}}, \bibinfo {author} {\bibfnamefont {Wonkeun}\ \bibnamefont
  {Chang}}, \bibinfo {author} {\bibfnamefont {John~C.}\ \bibnamefont
  {Travers}}, \ and\ \bibinfo {author} {\bibfnamefont {Philip~St.J.}\
  \bibnamefont {Russell}},\ }\bibfield  {title} {\enquote {\bibinfo {title}
  {Vacuum-ultraviolet to infrared supercontinuum in hydrogen-filled photonic
  crystal fiber},}\ }\href {\doibase 10.1364/OPTICA.2.000292} {\bibfield
  {journal} {\bibinfo  {journal} {Optica}\ }\textbf {\bibinfo {volume} {2}},\
  \bibinfo {pages} {292--300} (\bibinfo {year} {2015})}\BibitemShut {NoStop}%
\bibitem [{\citenamefont {Ermolov}\ \emph {et~al.}(2015)\citenamefont
  {Ermolov}, \citenamefont {Mak}, \citenamefont {Frosz}, \citenamefont
  {Travers},\ and\ \citenamefont {Russell}}]{Ermolov-2015-PhysRevA.92.033821}%
  \BibitemOpen
  \bibfield  {author} {\bibinfo {author} {\bibfnamefont {A.}~\bibnamefont
  {Ermolov}}, \bibinfo {author} {\bibfnamefont {K.~F.}\ \bibnamefont {Mak}},
  \bibinfo {author} {\bibfnamefont {M.~H.}\ \bibnamefont {Frosz}}, \bibinfo
  {author} {\bibfnamefont {J.~C.}\ \bibnamefont {Travers}}, \ and\ \bibinfo
  {author} {\bibfnamefont {P.~St.~J.}\ \bibnamefont {Russell}},\ }\bibfield
  {title} {\enquote {\bibinfo {title} {Supercontinuum generation in the vacuum
  ultraviolet through dispersive-wave and soliton-plasma interaction in a
  noble-gas-filled hollow-core photonic crystal fiber},}\ }\href {\doibase
  10.1103/PhysRevA.92.033821} {\bibfield  {journal} {\bibinfo  {journal} {Phys.
  Rev. A}\ }\textbf {\bibinfo {volume} {92}},\ \bibinfo {pages} {033821}
  (\bibinfo {year} {2015})}\BibitemShut {NoStop}%
\bibitem [{\citenamefont {Beaud}\ \emph {et~al.}(1987)\citenamefont {Beaud},
  \citenamefont {Hodel}, \citenamefont {Zysset},\ and\ \citenamefont
  {Weber}}]{beaud:1987}%
  \BibitemOpen
  \bibfield  {author} {\bibinfo {author} {\bibfnamefont {P.}~\bibnamefont
  {Beaud}}, \bibinfo {author} {\bibfnamefont {W.}~\bibnamefont {Hodel}},
  \bibinfo {author} {\bibfnamefont {B.}~\bibnamefont {Zysset}}, \ and\ \bibinfo
  {author} {\bibfnamefont {H.}~\bibnamefont {Weber}},\ }\bibfield  {title}
  {\enquote {\bibinfo {title} {Ultrashort pulse propagation, pulse breakup, and
  fundamental soliton formation in a single-mode optical fiber},}\ }\href
  {\doibase 10.1109/JQE.1987.1073262} {\bibfield  {journal} {\bibinfo
  {journal} {IEEE J. Quantum Electron.}\ }\textbf {\bibinfo {volume} {23}},\
  \bibinfo {pages} {1938 -- 1946} (\bibinfo {year} {1987})}\BibitemShut
  {NoStop}%
\bibitem [{\citenamefont {Wise}\ \emph {et~al.}(1988)\citenamefont {Wise},
  \citenamefont {Walmsley},\ and\ \citenamefont {Tang}}]{Wise:1988}%
  \BibitemOpen
  \bibfield  {author} {\bibinfo {author} {\bibfnamefont {F.~W.}\ \bibnamefont
  {Wise}}, \bibinfo {author} {\bibfnamefont {I.~A.}\ \bibnamefont {Walmsley}},
  \ and\ \bibinfo {author} {\bibfnamefont {C.~L.}\ \bibnamefont {Tang}},\
  }\bibfield  {title} {\enquote {\bibinfo {title} {Simultaneous formation of
  solitons and dispersive waves in a femtosecond ring dye laser},}\ }\href
  {\doibase 10.1364/OL.13.000129} {\bibfield  {journal} {\bibinfo  {journal}
  {Opt. Lett.}\ }\textbf {\bibinfo {volume} {13}},\ \bibinfo {pages} {129--131}
  (\bibinfo {year} {1988})}\BibitemShut {NoStop}%
\bibitem [{\citenamefont {Gouveia-Neto}\ \emph {et~al.}(1988)\citenamefont
  {Gouveia-Neto}, \citenamefont {Faldon},\ and\ \citenamefont
  {Taylor}}]{Gouveia-Neto:1988}%
  \BibitemOpen
  \bibfield  {author} {\bibinfo {author} {\bibfnamefont {A.~S.}\ \bibnamefont
  {Gouveia-Neto}}, \bibinfo {author} {\bibfnamefont {M.~E.}\ \bibnamefont
  {Faldon}}, \ and\ \bibinfo {author} {\bibfnamefont {J.~R.}\ \bibnamefont
  {Taylor}},\ }\bibfield  {title} {\enquote {\bibinfo {title} {Solitons in the
  region of the minimum group-velocity dispersion of single-mode optical
  fibers},}\ }\href {\doibase 10.1364/OL.13.000770} {\bibfield  {journal}
  {\bibinfo  {journal} {Opt. Lett.}\ }\textbf {\bibinfo {volume} {13}},\
  \bibinfo {pages} {770--772} (\bibinfo {year} {1988})}\BibitemShut {NoStop}%
\bibitem [{\citenamefont {Rubino}\ \emph
  {et~al.}(2012{\natexlab{a}})\citenamefont {Rubino}, \citenamefont
  {McLenaghan}, \citenamefont {Kehr}, \citenamefont {Belgiorno}, \citenamefont
  {Townsend}, \citenamefont {Rohr}, \citenamefont {Kuklewicz}, \citenamefont
  {Leonhardt}, \citenamefont {K\"onig},\ and\ \citenamefont
  {Faccio}}]{rubino:2012}%
  \BibitemOpen
  \bibfield  {author} {\bibinfo {author} {\bibfnamefont {E.}~\bibnamefont
  {Rubino}}, \bibinfo {author} {\bibfnamefont {J.}~\bibnamefont {McLenaghan}},
  \bibinfo {author} {\bibfnamefont {S.~C.}\ \bibnamefont {Kehr}}, \bibinfo
  {author} {\bibfnamefont {F.}~\bibnamefont {Belgiorno}}, \bibinfo {author}
  {\bibfnamefont {D.}~\bibnamefont {Townsend}}, \bibinfo {author}
  {\bibfnamefont {S.}~\bibnamefont {Rohr}}, \bibinfo {author} {\bibfnamefont
  {C.~E.}\ \bibnamefont {Kuklewicz}}, \bibinfo {author} {\bibfnamefont
  {U.}~\bibnamefont {Leonhardt}}, \bibinfo {author} {\bibfnamefont
  {F.}~\bibnamefont {K\"onig}}, \ and\ \bibinfo {author} {\bibfnamefont
  {D.}~\bibnamefont {Faccio}},\ }\bibfield  {title} {\enquote {\bibinfo {title}
  {Negative-frequency resonant radiation},}\ }\href {\doibase
  10.1103/PhysRevLett.108.253901} {\bibfield  {journal} {\bibinfo  {journal}
  {Phys. Rev. Lett.}\ }\textbf {\bibinfo {volume} {108}},\ \bibinfo {pages}
  {253901} (\bibinfo {year} {2012}{\natexlab{a}})}\BibitemShut {NoStop}%
\bibitem [{\citenamefont {Petev}\ \emph {et~al.}(2013)\citenamefont {Petev},
  \citenamefont {Westerberg}, \citenamefont {Moss}, \citenamefont {Rubino},
  \citenamefont {Rimoldi}, \citenamefont {Cacciatori}, \citenamefont
  {Belgiorno},\ and\ \citenamefont {Faccio}}]{petev.PhysRevLett.111.043902}%
  \BibitemOpen
  \bibfield  {author} {\bibinfo {author} {\bibfnamefont {M.}~\bibnamefont
  {Petev}}, \bibinfo {author} {\bibfnamefont {N.}~\bibnamefont {Westerberg}},
  \bibinfo {author} {\bibfnamefont {D.}~\bibnamefont {Moss}}, \bibinfo {author}
  {\bibfnamefont {E.}~\bibnamefont {Rubino}}, \bibinfo {author} {\bibfnamefont
  {C.}~\bibnamefont {Rimoldi}}, \bibinfo {author} {\bibfnamefont {S.~L.}\
  \bibnamefont {Cacciatori}}, \bibinfo {author} {\bibfnamefont
  {F.}~\bibnamefont {Belgiorno}}, \ and\ \bibinfo {author} {\bibfnamefont
  {D.}~\bibnamefont {Faccio}},\ }\bibfield  {title} {\enquote {\bibinfo {title}
  {Blackbody emission from light interacting with an effective moving
  dispersive medium},}\ }\href {\doibase 10.1103/PhysRevLett.111.043902}
  {\bibfield  {journal} {\bibinfo  {journal} {Phys. Rev. Lett.}\ }\textbf
  {\bibinfo {volume} {111}},\ \bibinfo {pages} {043902} (\bibinfo {year}
  {2013})}\BibitemShut {NoStop}%
\bibitem [{\citenamefont {Conforti}\ \emph
  {et~al.}(2013{\natexlab{a}})\citenamefont {Conforti}, \citenamefont
  {Westerberg}, \citenamefont {Baronio}, \citenamefont {Trillo},\ and\
  \citenamefont {Faccio}}]{conforti-PhysRevA.88.013829-2013}%
  \BibitemOpen
  \bibfield  {author} {\bibinfo {author} {\bibfnamefont {Matteo}\ \bibnamefont
  {Conforti}}, \bibinfo {author} {\bibfnamefont {Niclas}\ \bibnamefont
  {Westerberg}}, \bibinfo {author} {\bibfnamefont {Fabio}\ \bibnamefont
  {Baronio}}, \bibinfo {author} {\bibfnamefont {Stefano}\ \bibnamefont
  {Trillo}}, \ and\ \bibinfo {author} {\bibfnamefont {Daniele}\ \bibnamefont
  {Faccio}},\ }\bibfield  {title} {\enquote {\bibinfo {title}
  {Negative-frequency dispersive wave generation in quadratic media},}\ }\href
  {\doibase 10.1103/PhysRevA.88.013829} {\bibfield  {journal} {\bibinfo
  {journal} {Phys. Rev. A}\ }\textbf {\bibinfo {volume} {88}},\ \bibinfo
  {pages} {013829} (\bibinfo {year} {2013}{\natexlab{a}})}\BibitemShut
  {NoStop}%
\bibitem [{\citenamefont {Rubino}\ \emph
  {et~al.}(2012{\natexlab{b}})\citenamefont {Rubino}, \citenamefont {Lotti},
  \citenamefont {Belgiorno}, \citenamefont {Cacciatori}, \citenamefont
  {Couairon}, \citenamefont {Leonhardt},\ and\ \citenamefont
  {Faccio}}]{Rubino:2012-SciRep}%
  \BibitemOpen
  \bibfield  {author} {\bibinfo {author} {\bibfnamefont {E}~\bibnamefont
  {Rubino}}, \bibinfo {author} {\bibfnamefont {A}~\bibnamefont {Lotti}},
  \bibinfo {author} {\bibfnamefont {F}~\bibnamefont {Belgiorno}}, \bibinfo
  {author} {\bibfnamefont {S~L}\ \bibnamefont {Cacciatori}}, \bibinfo {author}
  {\bibfnamefont {A}~\bibnamefont {Couairon}}, \bibinfo {author} {\bibfnamefont
  {U}~\bibnamefont {Leonhardt}}, \ and\ \bibinfo {author} {\bibfnamefont
  {D}~\bibnamefont {Faccio}},\ }\bibfield  {title} {\enquote {\bibinfo {title}
  {{Soliton-induced relativistic-scattering and amplification.}}}\ }\href
  {\doibase 10.1038/srep00932} {\bibfield  {journal} {\bibinfo  {journal} {Sci.
  Rep.}\ }\textbf {\bibinfo {volume} {2}},\ \bibinfo {pages} {932} (\bibinfo
  {year} {2012}{\natexlab{b}})}\BibitemShut {NoStop}%
\bibitem [{\citenamefont {Conforti}\ \emph
  {et~al.}(2013{\natexlab{b}})\citenamefont {Conforti}, \citenamefont {Marini},
  \citenamefont {Tran}, \citenamefont {Faccio},\ and\ \citenamefont
  {Biancalana}}]{Conforti2013}%
  \BibitemOpen
  \bibfield  {author} {\bibinfo {author} {\bibfnamefont {Matteo}\ \bibnamefont
  {Conforti}}, \bibinfo {author} {\bibfnamefont {Andrea}\ \bibnamefont
  {Marini}}, \bibinfo {author} {\bibfnamefont {Truong~X.}\ \bibnamefont
  {Tran}}, \bibinfo {author} {\bibfnamefont {Daniele}\ \bibnamefont {Faccio}},
  \ and\ \bibinfo {author} {\bibfnamefont {Fabio}\ \bibnamefont {Biancalana}},\
  }\bibfield  {title} {\enquote {\bibinfo {title} {Interaction between optical
  fields and their conjugates in nonlinear media},}\ }\href {\doibase
  10.1364/OE.21.031239} {\bibfield  {journal} {\bibinfo  {journal} {Opt.
  Express}\ }\textbf {\bibinfo {volume} {21}},\ \bibinfo {pages} {31239--31252}
  (\bibinfo {year} {2013}{\natexlab{b}})}\BibitemShut {NoStop}%
\bibitem [{\citenamefont {Lour\'es}\ \emph {et~al.}(2015)\citenamefont
  {Lour\'es}, \citenamefont {Faccio},\ and\ \citenamefont
  {Biancalana}}]{loures-PhysRevLett.115.193904}%
  \BibitemOpen
  \bibfield  {author} {\bibinfo {author} {\bibfnamefont {Cristian~Redondo}\
  \bibnamefont {Lour\'es}}, \bibinfo {author} {\bibfnamefont {Daniele}\
  \bibnamefont {Faccio}}, \ and\ \bibinfo {author} {\bibfnamefont {Fabio}\
  \bibnamefont {Biancalana}},\ }\bibfield  {title} {\enquote {\bibinfo {title}
  {Nonlinear cavity and frequency comb radiations induced by negative frequency
  field effects},}\ }\href {\doibase 10.1103/PhysRevLett.115.193904} {\bibfield
   {journal} {\bibinfo  {journal} {Phys. Rev. Lett.}\ }\textbf {\bibinfo
  {volume} {115}},\ \bibinfo {pages} {193904} (\bibinfo {year}
  {2015})}\BibitemShut {NoStop}%
\bibitem [{\citenamefont {DeSalvo}\ \emph {et~al.}(1992)\citenamefont
  {DeSalvo}, \citenamefont {Hagan}, \citenamefont {{Sheik-Bahae}},
  \citenamefont {Stegeman}, \citenamefont {{Van Stryland}},\ and\ \citenamefont
  {Vanherzeele}}]{desalvo:1992}%
  \BibitemOpen
  \bibfield  {author} {\bibinfo {author} {\bibfnamefont {R.}~\bibnamefont
  {DeSalvo}}, \bibinfo {author} {\bibfnamefont {D.J.}\ \bibnamefont {Hagan}},
  \bibinfo {author} {\bibfnamefont {M.}~\bibnamefont {{Sheik-Bahae}}}, \bibinfo
  {author} {\bibfnamefont {G.}~\bibnamefont {Stegeman}}, \bibinfo {author}
  {\bibfnamefont {E.~W.}\ \bibnamefont {{Van Stryland}}}, \ and\ \bibinfo
  {author} {\bibfnamefont {H.}~\bibnamefont {Vanherzeele}},\ }\bibfield
  {title} {\enquote {\bibinfo {title} {Self-focusing and self-defocusing by
  cascaded second-order effects in {KTP}},}\ }\href {\doibase
  10.1364/OL.17.000028} {\bibfield  {journal} {\bibinfo  {journal} {Opt.
  Lett.}\ }\textbf {\bibinfo {volume} {17}},\ \bibinfo {pages} {28--30}
  (\bibinfo {year} {1992})}\BibitemShut {NoStop}%
\bibitem [{\citenamefont {Liu}\ \emph {et~al.}(1999)\citenamefont {Liu},
  \citenamefont {Qian},\ and\ \citenamefont {Wise}}]{liu:1999}%
  \BibitemOpen
  \bibfield  {author} {\bibinfo {author} {\bibfnamefont {X.}~\bibnamefont
  {Liu}}, \bibinfo {author} {\bibfnamefont {L.-J.}\ \bibnamefont {Qian}}, \
  and\ \bibinfo {author} {\bibfnamefont {F.~W.}\ \bibnamefont {Wise}},\
  }\bibfield  {title} {\enquote {\bibinfo {title} {High-energy pulse
  compression by use of negative phase shifts produced by the cascaded
  $\chi^{(2)}:\chi^{(2)}$ nonlinearity},}\ }\href {\doibase
  10.1364/OL.24.001777} {\bibfield  {journal} {\bibinfo  {journal} {Opt.
  Lett.}\ }\textbf {\bibinfo {volume} {24}},\ \bibinfo {pages} {1777--1779}
  (\bibinfo {year} {1999})}\BibitemShut {NoStop}%
\bibitem [{\citenamefont {Ashihara}\ \emph {et~al.}(2002)\citenamefont
  {Ashihara}, \citenamefont {Nishina}, \citenamefont {Shimura},\ and\
  \citenamefont {Kuroda}}]{ashihara:2002}%
  \BibitemOpen
  \bibfield  {author} {\bibinfo {author} {\bibfnamefont {S.}~\bibnamefont
  {Ashihara}}, \bibinfo {author} {\bibfnamefont {J.}~\bibnamefont {Nishina}},
  \bibinfo {author} {\bibfnamefont {T.}~\bibnamefont {Shimura}}, \ and\
  \bibinfo {author} {\bibfnamefont {K.}~\bibnamefont {Kuroda}},\ }\bibfield
  {title} {\enquote {\bibinfo {title} {Soliton compression of femtosecond
  pulses in quadratic media},}\ }\href {\doibase 10.1364/JOSAB.19.002505}
  {\bibfield  {journal} {\bibinfo  {journal} {J. Opt. Soc. Am. B}\ }\textbf
  {\bibinfo {volume} {19}},\ \bibinfo {pages} {2505--2510} (\bibinfo {year}
  {2002})}\BibitemShut {NoStop}%
\bibitem [{\citenamefont {Langrock}\ \emph {et~al.}(2007)\citenamefont
  {Langrock}, \citenamefont {Fejer}, \citenamefont {Hartl},\ and\ \citenamefont
  {Fermann}}]{Langrock:2007}%
  \BibitemOpen
  \bibfield  {author} {\bibinfo {author} {\bibfnamefont {Carsten}\ \bibnamefont
  {Langrock}}, \bibinfo {author} {\bibfnamefont {M.~M.}\ \bibnamefont {Fejer}},
  \bibinfo {author} {\bibfnamefont {I.}~\bibnamefont {Hartl}}, \ and\ \bibinfo
  {author} {\bibfnamefont {Martin~E.}\ \bibnamefont {Fermann}},\ }\bibfield
  {title} {\enquote {\bibinfo {title} {Generation of octave-spanning spectra
  inside reverse-proton-exchanged periodically poled lithium niobate
  waveguides},}\ }\href {\doibase 10.1364/OL.32.002478} {\bibfield  {journal}
  {\bibinfo  {journal} {Opt. Lett.}\ }\textbf {\bibinfo {volume} {32}},\
  \bibinfo {pages} {2478--2480} (\bibinfo {year} {2007})}\BibitemShut {NoStop}%
\bibitem [{\citenamefont {Phillips}\ \emph
  {et~al.}(2011{\natexlab{a}})\citenamefont {Phillips}, \citenamefont
  {Langrock}, \citenamefont {Pelc}, \citenamefont {Fejer}, \citenamefont
  {Jiang}, \citenamefont {Fermann},\ and\ \citenamefont
  {Hartl}}]{Phillips:2011-ol}%
  \BibitemOpen
  \bibfield  {author} {\bibinfo {author} {\bibfnamefont {C.~R.}\ \bibnamefont
  {Phillips}}, \bibinfo {author} {\bibfnamefont {Carsten}\ \bibnamefont
  {Langrock}}, \bibinfo {author} {\bibfnamefont {J.~S.}\ \bibnamefont {Pelc}},
  \bibinfo {author} {\bibfnamefont {M.~M.}\ \bibnamefont {Fejer}}, \bibinfo
  {author} {\bibfnamefont {J.}~\bibnamefont {Jiang}}, \bibinfo {author}
  {\bibfnamefont {Martin~E.}\ \bibnamefont {Fermann}}, \ and\ \bibinfo {author}
  {\bibfnamefont {I.}~\bibnamefont {Hartl}},\ }\bibfield  {title} {\enquote
  {\bibinfo {title} {Supercontinuum generation in quasi-phase-matched
  {LiNbO}$_3$ waveguide pumped by a {T}m-doped fiber laser system},}\ }\href
  {\doibase 10.1364/OL.36.003912} {\bibfield  {journal} {\bibinfo  {journal}
  {Opt. Lett.}\ }\textbf {\bibinfo {volume} {36}},\ \bibinfo {pages}
  {3912--3914} (\bibinfo {year} {2011}{\natexlab{a}})}\BibitemShut {NoStop}%
\bibitem [{\citenamefont {Phillips}\ \emph
  {et~al.}(2011{\natexlab{b}})\citenamefont {Phillips}, \citenamefont
  {Langrock}, \citenamefont {Pelc}, \citenamefont {Fejer}, \citenamefont
  {Hartl},\ and\ \citenamefont {Fermann}}]{Phillips:2011}%
  \BibitemOpen
  \bibfield  {author} {\bibinfo {author} {\bibfnamefont {C.~R.}\ \bibnamefont
  {Phillips}}, \bibinfo {author} {\bibfnamefont {Carsten}\ \bibnamefont
  {Langrock}}, \bibinfo {author} {\bibfnamefont {J.~S.}\ \bibnamefont {Pelc}},
  \bibinfo {author} {\bibfnamefont {M.~M.}\ \bibnamefont {Fejer}}, \bibinfo
  {author} {\bibfnamefont {I.}~\bibnamefont {Hartl}}, \ and\ \bibinfo {author}
  {\bibfnamefont {Martin~E.}\ \bibnamefont {Fermann}},\ }\bibfield  {title}
  {\enquote {\bibinfo {title} {Supercontinuum generation in quasi-phasematched
  waveguides},}\ }\href {\doibase 10.1364/OE.19.018754} {\bibfield  {journal}
  {\bibinfo  {journal} {Opt. Express}\ }\textbf {\bibinfo {volume} {19}},\
  \bibinfo {pages} {18754--18773} (\bibinfo {year}
  {2011}{\natexlab{b}})}\BibitemShut {NoStop}%
\bibitem [{\citenamefont {Guo}\ \emph {et~al.}(2015)\citenamefont {Guo},
  \citenamefont {Zhou}, \citenamefont {Steinert}, \citenamefont {Setzpfandt},
  \citenamefont {Pertsch}, \citenamefont {Chung}, \citenamefont {Chen},\ and\
  \citenamefont {Bache}}]{Guo-LN-exp-2015}%
  \BibitemOpen
  \bibfield  {author} {\bibinfo {author} {\bibfnamefont {H.~R.}\ \bibnamefont
  {Guo}}, \bibinfo {author} {\bibfnamefont {B.~B.}\ \bibnamefont {Zhou}},
  \bibinfo {author} {\bibfnamefont {M.}~\bibnamefont {Steinert}}, \bibinfo
  {author} {\bibfnamefont {F.}~\bibnamefont {Setzpfandt}}, \bibinfo {author}
  {\bibfnamefont {T.}~\bibnamefont {Pertsch}}, \bibinfo {author} {\bibfnamefont
  {H.~P.}\ \bibnamefont {Chung}}, \bibinfo {author} {\bibfnamefont {Y.~H.}\
  \bibnamefont {Chen}}, \ and\ \bibinfo {author} {\bibfnamefont
  {M.}~\bibnamefont {Bache}},\ }\bibfield  {title} {\enquote {\bibinfo {title}
  {Supercontinuum generation in quadratic nonlinear waveguides without
  quasi-phase matching},}\ }\href {\doibase 10.1364/OL.40.000629} {\bibfield
  {journal} {\bibinfo  {journal} {Opt. Lett.}\ }\textbf {\bibinfo {volume}
  {40}},\ \bibinfo {pages} {629--632} (\bibinfo {year} {2015})}\BibitemShut
  {NoStop}%
\bibitem [{\citenamefont {Zhou}\ \emph {et~al.}(2012)\citenamefont {Zhou},
  \citenamefont {Chong}, \citenamefont {Wise},\ and\ \citenamefont
  {Bache}}]{zhou:2012}%
  \BibitemOpen
  \bibfield  {author} {\bibinfo {author} {\bibfnamefont {B.~B.}\ \bibnamefont
  {Zhou}}, \bibinfo {author} {\bibfnamefont {A.}~\bibnamefont {Chong}},
  \bibinfo {author} {\bibfnamefont {F.~W.}\ \bibnamefont {Wise}}, \ and\
  \bibinfo {author} {\bibfnamefont {M.}~\bibnamefont {Bache}},\ }\bibfield
  {title} {\enquote {\bibinfo {title} {Ultrafast and octave-spanning optical
  nonlinearities from strongly phase-mismatched quadratic interactions},}\
  }\href {\doibase 10.1103/PhysRevLett.109.043902} {\bibfield  {journal}
  {\bibinfo  {journal} {Phys. Rev. Lett.}\ }\textbf {\bibinfo {volume} {109}},\
  \bibinfo {pages} {043902} (\bibinfo {year} {2012})}\BibitemShut {NoStop}%
\bibitem [{\citenamefont {Zhou}\ \emph {et~al.}(2015)\citenamefont {Zhou},
  \citenamefont {Guo},\ and\ \citenamefont {Bache}}]{Zhou:2015}%
  \BibitemOpen
  \bibfield  {author} {\bibinfo {author} {\bibfnamefont {Binbin}\ \bibnamefont
  {Zhou}}, \bibinfo {author} {\bibfnamefont {Hairun}\ \bibnamefont {Guo}}, \
  and\ \bibinfo {author} {\bibfnamefont {Morten}\ \bibnamefont {Bache}},\
  }\bibfield  {title} {\enquote {\bibinfo {title} {Energetic mid-{IR}
  femtosecond pulse generation by self-defocusing soliton-induced dispersive
  waves in a bulk quadratic nonlinear crystal},}\ }\href {\doibase
  10.1364/OE.23.006924} {\bibfield  {journal} {\bibinfo  {journal} {Opt.
  Express}\ }\textbf {\bibinfo {volume} {23}},\ \bibinfo {pages} {6924--6936}
  (\bibinfo {year} {2015})}\BibitemShut {NoStop}%
\bibitem [{\citenamefont {Zhou}\ and\ \citenamefont
  {Bache}(2015)}]{Zhou:2015-OL}%
  \BibitemOpen
  \bibfield  {author} {\bibinfo {author} {\bibfnamefont {Binbin}\ \bibnamefont
  {Zhou}}\ and\ \bibinfo {author} {\bibfnamefont {Morten}\ \bibnamefont
  {Bache}},\ }\bibfield  {title} {\enquote {\bibinfo {title} {Dispersive waves
  induced by self-defocusing temporal solitons in a beta-barium-borate
  crystal},}\ }\href {\doibase 10.1364/OL.40.004257} {\bibfield  {journal}
  {\bibinfo  {journal} {Opt. Lett.}\ }\textbf {\bibinfo {volume} {40}},\
  \bibinfo {pages} {4257--4260} (\bibinfo {year} {2015})}\BibitemShut {NoStop}%
\bibitem [{\citenamefont {Zhou}\ and\ \citenamefont
  {Bache}(2016)}]{Zhou2016-aplp}%
  \BibitemOpen
  \bibfield  {author} {\bibinfo {author} {\bibfnamefont {Binbin}\ \bibnamefont
  {Zhou}}\ and\ \bibinfo {author} {\bibfnamefont {Morten}\ \bibnamefont
  {Bache}},\ }\bibfield  {title} {\enquote {\bibinfo {title} {Invited article:
  Multiple-octave spanning high-energy mid-{IR} supercontinuum generation in
  bulk quadratic nonlinear crystals},}\ }\href {\doibase 10.1063/1.4953177}
  {\bibfield  {journal} {\bibinfo  {journal} {APL Photonics}\ }\textbf
  {\bibinfo {volume} {1}},\ \bibinfo {eid} {050802} (\bibinfo {year}
  {2016})}\BibitemShut {NoStop}%
\bibitem [{\citenamefont {Conforti}\ \emph
  {et~al.}(2010{\natexlab{a}})\citenamefont {Conforti}, \citenamefont
  {Baronio},\ and\ \citenamefont {De~Angelis}}]{conforti:2010PRA}%
  \BibitemOpen
  \bibfield  {author} {\bibinfo {author} {\bibfnamefont {Matteo}\ \bibnamefont
  {Conforti}}, \bibinfo {author} {\bibfnamefont {Fabio}\ \bibnamefont
  {Baronio}}, \ and\ \bibinfo {author} {\bibfnamefont {Costantino}\
  \bibnamefont {De~Angelis}},\ }\bibfield  {title} {\enquote {\bibinfo {title}
  {Nonlinear envelope equation for broadband optical pulses in quadratic
  media},}\ }\href {\doibase 10.1103/PhysRevA.81.053841} {\bibfield  {journal}
  {\bibinfo  {journal} {Phys. Rev. A}\ }\textbf {\bibinfo {volume} {81}},\
  \bibinfo {pages} {053841} (\bibinfo {year} {2010}{\natexlab{a}})}\BibitemShut
  {NoStop}%
\bibitem [{\citenamefont {Bache}(2016)}]{bache:2016-NAEE}%
  \BibitemOpen
  \bibfield  {author} {\bibinfo {author} {\bibfnamefont {Morten}\ \bibnamefont
  {Bache}},\ }\href@noop {} {\enquote {\bibinfo {title} {The nonlinear
  analytical envelope equation in quadratic nonlinear crystals},}\ } (\bibinfo
  {year} {2016}),\ \bibinfo {note} {{arXiv:1603.00188}}\BibitemShut {NoStop}%
\bibitem [{\citenamefont {Baronio}\ \emph {et~al.}(2012)\citenamefont
  {Baronio}, \citenamefont {Conforti}, \citenamefont {Angelis}, \citenamefont
  {Modotto}, \citenamefont {Wabnitz}, \citenamefont {Andreana}, \citenamefont
  {Tonello}, \citenamefont {Leproux},\ and\ \citenamefont
  {Couderc}}]{Baronio:2012}%
  \BibitemOpen
  \bibfield  {author} {\bibinfo {author} {\bibfnamefont {Fabio}\ \bibnamefont
  {Baronio}}, \bibinfo {author} {\bibfnamefont {Matteo}\ \bibnamefont
  {Conforti}}, \bibinfo {author} {\bibfnamefont {Costantino~De}\ \bibnamefont
  {Angelis}}, \bibinfo {author} {\bibfnamefont {Daniele}\ \bibnamefont
  {Modotto}}, \bibinfo {author} {\bibfnamefont {Stefan}\ \bibnamefont
  {Wabnitz}}, \bibinfo {author} {\bibfnamefont {Marco}\ \bibnamefont
  {Andreana}}, \bibinfo {author} {\bibfnamefont {Alessandro}\ \bibnamefont
  {Tonello}}, \bibinfo {author} {\bibfnamefont {Philippe}\ \bibnamefont
  {Leproux}}, \ and\ \bibinfo {author} {\bibfnamefont {Vincent}\ \bibnamefont
  {Couderc}},\ }\bibfield  {title} {\enquote {\bibinfo {title} {Second and
  third order susceptibilities mixing for supercontinuum generation and
  shaping},}\ }\href {\doibase 10.1016/j.yofte.2012.07.001} {\bibfield
  {journal} {\bibinfo  {journal} {Opt. Fiber Technol.}\ }\textbf {\bibinfo
  {volume} {18}},\ \bibinfo {pages} {283 -- 289} (\bibinfo {year}
  {2012})}\BibitemShut {NoStop}%
\bibitem [{PRL()}]{PRL-2017-suppl}%
  \BibitemOpen
  \href@noop {} {}\bibinfo {note} {See the Appendix for Supplemental Material,
  which includes Refs.
  \cite{conforti:2010,Guo:2013,gayer:2008,Lee2000,Lhuillier2007a,Lhuillier2007,Shoji:1997},
  for a complete description of the envelope equation, the derivation of the
  phase matching conditions and discussion of how quasiphase matching is
  introduced, and additional numerical simulation examples.}\BibitemShut
  {Stop}%
\bibitem [{\citenamefont {Skryabin}\ and\ \citenamefont
  {Yulin}(2005)}]{skryabin:2005}%
  \BibitemOpen
  \bibfield  {author} {\bibinfo {author} {\bibfnamefont {D.~V.}\ \bibnamefont
  {Skryabin}}\ and\ \bibinfo {author} {\bibfnamefont {A.~V.}\ \bibnamefont
  {Yulin}},\ }\bibfield  {title} {\enquote {\bibinfo {title} {Theory of
  generation of new frequencies by mixing of solitons and dispersive waves in
  optical fibers},}\ }\href {\doibase 10.1103/PhysRevE.72.016619} {\bibfield
  {journal} {\bibinfo  {journal} {Phys. Rev. E}\ }\textbf {\bibinfo {volume}
  {72}},\ \bibinfo {pages} {016619} (\bibinfo {year} {2005})}\BibitemShut
  {NoStop}%
\bibitem [{\citenamefont {Bache}\ \emph {et~al.}(2010)\citenamefont {Bache},
  \citenamefont {Bang}, \citenamefont {Zhou}, \citenamefont {Moses},\ and\
  \citenamefont {Wise}}]{bache:2010e}%
  \BibitemOpen
  \bibfield  {author} {\bibinfo {author} {\bibfnamefont {M.}~\bibnamefont
  {Bache}}, \bibinfo {author} {\bibfnamefont {O.}~\bibnamefont {Bang}},
  \bibinfo {author} {\bibfnamefont {B.~B.}\ \bibnamefont {Zhou}}, \bibinfo
  {author} {\bibfnamefont {J.}~\bibnamefont {Moses}}, \ and\ \bibinfo {author}
  {\bibfnamefont {F.~W.}\ \bibnamefont {Wise}},\ }\bibfield  {title} {\enquote
  {\bibinfo {title} {Optical {C}herenkov radiation in ultrafast cascaded
  second-harmonic generation},}\ }\href {\doibase 10.1103/PhysRevA.82.063806}
  {\bibfield  {journal} {\bibinfo  {journal} {Phys. Rev. A}\ }\textbf {\bibinfo
  {volume} {82}},\ \bibinfo {pages} {063806} (\bibinfo {year}
  {2010})}\BibitemShut {NoStop}%
\bibitem [{\citenamefont {Bache}\ \emph
  {et~al.}(2007{\natexlab{a}})\citenamefont {Bache}, \citenamefont {Bang},
  \citenamefont {Moses},\ and\ \citenamefont {Wise}}]{bache:2007a}%
  \BibitemOpen
  \bibfield  {author} {\bibinfo {author} {\bibfnamefont {M.}~\bibnamefont
  {Bache}}, \bibinfo {author} {\bibfnamefont {O.}~\bibnamefont {Bang}},
  \bibinfo {author} {\bibfnamefont {J.}~\bibnamefont {Moses}}, \ and\ \bibinfo
  {author} {\bibfnamefont {Frank~W.}\ \bibnamefont {Wise}},\ }\bibfield
  {title} {\enquote {\bibinfo {title} {Nonlocal explanation of stationary and
  nonstationary regimes in cascaded soliton pulse compression},}\ }\href
  {\doibase 10.1364/OL.32.002490} {\bibfield  {journal} {\bibinfo  {journal}
  {Opt. Lett.}\ }\textbf {\bibinfo {volume} {32}},\ \bibinfo {pages}
  {2490--2492} (\bibinfo {year} {2007}{\natexlab{a}})}\BibitemShut {NoStop}%
\bibitem [{\citenamefont {Bache}\ \emph {et~al.}(2008)\citenamefont {Bache},
  \citenamefont {Bang}, \citenamefont {Krolikowski}, \citenamefont {Moses},\
  and\ \citenamefont {Wise}}]{bache:2008}%
  \BibitemOpen
  \bibfield  {author} {\bibinfo {author} {\bibfnamefont {M.}~\bibnamefont
  {Bache}}, \bibinfo {author} {\bibfnamefont {O.}~\bibnamefont {Bang}},
  \bibinfo {author} {\bibfnamefont {W.}~\bibnamefont {Krolikowski}}, \bibinfo
  {author} {\bibfnamefont {J.}~\bibnamefont {Moses}}, \ and\ \bibinfo {author}
  {\bibfnamefont {Frank~W.}\ \bibnamefont {Wise}},\ }\bibfield  {title}
  {\enquote {\bibinfo {title} {Limits to compression with cascaded quadratic
  soliton compressors},}\ }\href {\doibase 10.1364/OE.16.003273} {\bibfield
  {journal} {\bibinfo  {journal} {Opt. Express}\ }\textbf {\bibinfo {volume}
  {16}},\ \bibinfo {pages} {3273--3287} (\bibinfo {year} {2008})}\BibitemShut
  {NoStop}%
\bibitem [{\citenamefont {Zhou}\ \emph {et~al.}(2014)\citenamefont {Zhou},
  \citenamefont {Guo},\ and\ \citenamefont {Bache}}]{zhou:PhysRevA.90.013823}%
  \BibitemOpen
  \bibfield  {author} {\bibinfo {author} {\bibfnamefont {Binbin}\ \bibnamefont
  {Zhou}}, \bibinfo {author} {\bibfnamefont {Hairun}\ \bibnamefont {Guo}}, \
  and\ \bibinfo {author} {\bibfnamefont {Morten}\ \bibnamefont {Bache}},\
  }\bibfield  {title} {\enquote {\bibinfo {title} {Soliton-induced nonlocal
  resonances observed through high-intensity tunable spectrally compressed
  second-harmonic peaks},}\ }\href {\doibase 10.1103/PhysRevA.90.013823}
  {\bibfield  {journal} {\bibinfo  {journal} {Phys. Rev. A}\ }\textbf {\bibinfo
  {volume} {90}},\ \bibinfo {pages} {013823} (\bibinfo {year}
  {2014})}\BibitemShut {NoStop}%
\bibitem [{\citenamefont {Nahata}\ \emph {et~al.}(1996)\citenamefont {Nahata},
  \citenamefont {Weling},\ and\ \citenamefont {Heinz}}]{Nahata1996}%
  \BibitemOpen
  \bibfield  {author} {\bibinfo {author} {\bibfnamefont {Ajay}\ \bibnamefont
  {Nahata}}, \bibinfo {author} {\bibfnamefont {Aniruddha~S.}\ \bibnamefont
  {Weling}}, \ and\ \bibinfo {author} {\bibfnamefont {Tony~F.}\ \bibnamefont
  {Heinz}},\ }\bibfield  {title} {\enquote {\bibinfo {title} {A wideband
  coherent terahertz spectroscopy system using optical rectification and
  electro-optic sampling},}\ }\href {\doibase
  http://dx.doi.org/10.1063/1.117511} {\bibfield  {journal} {\bibinfo
  {journal} {Appl. Phys. Lett.}\ }\textbf {\bibinfo {volume} {69}},\ \bibinfo
  {pages} {2321--2323} (\bibinfo {year} {1996})}\BibitemShut {NoStop}%
\bibitem [{\citenamefont {Bache}\ \emph
  {et~al.}(2007{\natexlab{b}})\citenamefont {Bache}, \citenamefont {Moses},\
  and\ \citenamefont {Wise}}]{bache:2007}%
  \BibitemOpen
  \bibfield  {author} {\bibinfo {author} {\bibfnamefont {M.}~\bibnamefont
  {Bache}}, \bibinfo {author} {\bibfnamefont {J.}~\bibnamefont {Moses}}, \ and\
  \bibinfo {author} {\bibfnamefont {F.~W.}\ \bibnamefont {Wise}},\ }\bibfield
  {title} {\enquote {\bibinfo {title} {Scaling laws for soliton pulse
  compression by cascaded quadratic nonlinearities},}\ }\href {\doibase
  10.1364/JOSAB.24.002752} {\bibfield  {journal} {\bibinfo  {journal} {J. Opt.
  Soc. Am. B}\ }\textbf {\bibinfo {volume} {24}},\ \bibinfo {pages}
  {2752--2762} (\bibinfo {year} {2007}{\natexlab{b}})},\ \bibinfo {note}
  {[erratum: {\it ibid.}, {\bf 27}, 2505 (2010)]}\BibitemShut {NoStop}%
\bibitem [{\citenamefont {Valiulis}\ \emph {et~al.}(2011)\citenamefont
  {Valiulis}, \citenamefont {Jukna}, \citenamefont {Jedrkiewicz}, \citenamefont
  {Clerici}, \citenamefont {Rubino},\ and\ \citenamefont
  {DiTrapani}}]{Valiulis:2011}%
  \BibitemOpen
  \bibfield  {author} {\bibinfo {author} {\bibfnamefont {G.}~\bibnamefont
  {Valiulis}}, \bibinfo {author} {\bibfnamefont {V.}~\bibnamefont {Jukna}},
  \bibinfo {author} {\bibfnamefont {O.}~\bibnamefont {Jedrkiewicz}}, \bibinfo
  {author} {\bibfnamefont {M.}~\bibnamefont {Clerici}}, \bibinfo {author}
  {\bibfnamefont {E.}~\bibnamefont {Rubino}}, \ and\ \bibinfo {author}
  {\bibfnamefont {P.}~\bibnamefont {DiTrapani}},\ }\bibfield  {title} {\enquote
  {\bibinfo {title} {Propagation dynamics and {X}-pulse formation in
  phase-mismatched second-harmonic generation},}\ }\href {\doibase
  10.1103/PhysRevA.83.043834} {\bibfield  {journal} {\bibinfo  {journal} {Phys.
  Rev. A}\ }\textbf {\bibinfo {volume} {83}},\ \bibinfo {pages} {043834}
  (\bibinfo {year} {2011})}\BibitemShut {NoStop}%
\bibitem [{\citenamefont {Del'Haye}\ \emph {et~al.}(2007)\citenamefont
  {Del'Haye}, \citenamefont {Schliesser}, \citenamefont {Arcizet},
  \citenamefont {Wilken}, \citenamefont {Holzwarth},\ and\ \citenamefont
  {Kippenberg}}]{DelHaye2007}%
  \BibitemOpen
  \bibfield  {author} {\bibinfo {author} {\bibfnamefont {P.}~\bibnamefont
  {Del'Haye}}, \bibinfo {author} {\bibfnamefont {A.}~\bibnamefont
  {Schliesser}}, \bibinfo {author} {\bibfnamefont {O.}~\bibnamefont {Arcizet}},
  \bibinfo {author} {\bibfnamefont {T.}~\bibnamefont {Wilken}}, \bibinfo
  {author} {\bibfnamefont {R.}~\bibnamefont {Holzwarth}}, \ and\ \bibinfo
  {author} {\bibfnamefont {T.~J.}\ \bibnamefont {Kippenberg}},\ }\bibfield
  {title} {\enquote {\bibinfo {title} {Optical frequency comb generation from a
  monolithic microresonator},}\ }\href {\doibase 10.1038/nature06401}
  {\bibfield  {journal} {\bibinfo  {journal} {Nature}\ }\textbf {\bibinfo
  {volume} {450}},\ \bibinfo {pages} {1214--1217} (\bibinfo {year}
  {2007})}\BibitemShut {NoStop}%
\bibitem [{\citenamefont {Herr}\ \emph {et~al.}(2014)\citenamefont {Herr},
  \citenamefont {Brasch}, \citenamefont {Jost}, \citenamefont {Wang},
  \citenamefont {Kondratiev}, \citenamefont {Gorodetsky},\ and\ \citenamefont
  {Kippenberg}}]{Herr.2014}%
  \BibitemOpen
  \bibfield  {author} {\bibinfo {author} {\bibfnamefont {T.}~\bibnamefont
  {Herr}}, \bibinfo {author} {\bibfnamefont {V.}~\bibnamefont {Brasch}},
  \bibinfo {author} {\bibfnamefont {J.~D.}\ \bibnamefont {Jost}}, \bibinfo
  {author} {\bibfnamefont {Y.~C.}\ \bibnamefont {Wang}}, \bibinfo {author}
  {\bibfnamefont {M.~N.}\ \bibnamefont {Kondratiev}}, \bibinfo {author}
  {\bibfnamefont {L.~M.}\ \bibnamefont {Gorodetsky}}, \ and\ \bibinfo {author}
  {\bibfnamefont {T.~J.}\ \bibnamefont {Kippenberg}},\ }\bibfield  {title}
  {\enquote {\bibinfo {title} {Temporal solitons in optical microresonators},}\
  }\href {\doibase 10.1038/nphoton.2013.343} {\bibfield  {journal} {\bibinfo
  {journal} {Nat. Photon.}\ }\textbf {\bibinfo {volume} {8}},\ \bibinfo {pages}
  {145--152} (\bibinfo {year} {2014})}\BibitemShut {NoStop}%
\bibitem [{\citenamefont {Brasch}\ \emph {et~al.}(2016)\citenamefont {Brasch},
  \citenamefont {Geiselmann}, \citenamefont {Herr}, \citenamefont {Lihachev},
  \citenamefont {Pfeiffer}, \citenamefont {Gorodetsky},\ and\ \citenamefont
  {Kippenberg}}]{Brasch2016}%
  \BibitemOpen
  \bibfield  {author} {\bibinfo {author} {\bibfnamefont {V.}~\bibnamefont
  {Brasch}}, \bibinfo {author} {\bibfnamefont {M.}~\bibnamefont {Geiselmann}},
  \bibinfo {author} {\bibfnamefont {T.}~\bibnamefont {Herr}}, \bibinfo {author}
  {\bibfnamefont {G.}~\bibnamefont {Lihachev}}, \bibinfo {author}
  {\bibfnamefont {M.~H.~P.}\ \bibnamefont {Pfeiffer}}, \bibinfo {author}
  {\bibfnamefont {M.~L.}\ \bibnamefont {Gorodetsky}}, \ and\ \bibinfo {author}
  {\bibfnamefont {T.~J.}\ \bibnamefont {Kippenberg}},\ }\bibfield  {title}
  {\enquote {\bibinfo {title} {Photonic chip{\textendash}based optical
  frequency comb using soliton {C}herenkov radiation},}\ }\href {\doibase
  10.1126/science.aad4811} {\bibfield  {journal} {\bibinfo  {journal}
  {Science}\ }\textbf {\bibinfo {volume} {351}},\ \bibinfo {pages} {357--360}
  (\bibinfo {year} {2016})}\BibitemShut {NoStop}%
\bibitem [{\citenamefont {Jung}\ \emph {et~al.}(2014)\citenamefont {Jung},
  \citenamefont {Stoll}, \citenamefont {Guo}, \citenamefont {Fischer},\ and\
  \citenamefont {Tang}}]{Jung2014}%
  \BibitemOpen
  \bibfield  {author} {\bibinfo {author} {\bibfnamefont {Hojoong}\ \bibnamefont
  {Jung}}, \bibinfo {author} {\bibfnamefont {Rebecca}\ \bibnamefont {Stoll}},
  \bibinfo {author} {\bibfnamefont {Xiang}\ \bibnamefont {Guo}}, \bibinfo
  {author} {\bibfnamefont {Debra}\ \bibnamefont {Fischer}}, \ and\ \bibinfo
  {author} {\bibfnamefont {Hong~X.}\ \bibnamefont {Tang}},\ }\bibfield  {title}
  {\enquote {\bibinfo {title} {Green, red, and {IR} frequency comb line
  generation from single {IR} pump in {AlN} microring resonator},}\ }\href
  {\doibase 10.1364/OPTICA.1.000396} {\bibfield  {journal} {\bibinfo  {journal}
  {Optica}\ }\textbf {\bibinfo {volume} {1}},\ \bibinfo {pages} {396--399}
  (\bibinfo {year} {2014})}\BibitemShut {NoStop}%
\bibitem [{\citenamefont {Miller}\ \emph {et~al.}(2014)\citenamefont {Miller},
  \citenamefont {Luke}, \citenamefont {Okawachi}, \citenamefont {Cardenas},
  \citenamefont {Gaeta},\ and\ \citenamefont {Lipson}}]{Miller2014}%
  \BibitemOpen
  \bibfield  {author} {\bibinfo {author} {\bibfnamefont {Steven}\ \bibnamefont
  {Miller}}, \bibinfo {author} {\bibfnamefont {Kevin}\ \bibnamefont {Luke}},
  \bibinfo {author} {\bibfnamefont {Yoshitomo}\ \bibnamefont {Okawachi}},
  \bibinfo {author} {\bibfnamefont {Jaime}\ \bibnamefont {Cardenas}}, \bibinfo
  {author} {\bibfnamefont {Alexander~L.}\ \bibnamefont {Gaeta}}, \ and\
  \bibinfo {author} {\bibfnamefont {Michal}\ \bibnamefont {Lipson}},\
  }\bibfield  {title} {\enquote {\bibinfo {title} {On-chip frequency comb
  generation at visible wavelengths via simultaneous second- and third-order
  optical nonlinearities},}\ }\href {\doibase 10.1364/OE.22.026517} {\bibfield
  {journal} {\bibinfo  {journal} {Opt. Express}\ }\textbf {\bibinfo {volume}
  {22}},\ \bibinfo {pages} {26517--26525} (\bibinfo {year} {2014})}\BibitemShut
  {NoStop}%
\bibitem [{\citenamefont {Ricciardi}\ \emph {et~al.}(2015)\citenamefont
  {Ricciardi}, \citenamefont {Mosca}, \citenamefont {Parisi}, \citenamefont
  {Maddaloni}, \citenamefont {Santamaria}, \citenamefont {De~Natale},\ and\
  \citenamefont {De~Rosa}}]{Ricciardi-PhysRevA.91.063839}%
  \BibitemOpen
  \bibfield  {author} {\bibinfo {author} {\bibfnamefont {Iolanda}\ \bibnamefont
  {Ricciardi}}, \bibinfo {author} {\bibfnamefont {Simona}\ \bibnamefont
  {Mosca}}, \bibinfo {author} {\bibfnamefont {Maria}\ \bibnamefont {Parisi}},
  \bibinfo {author} {\bibfnamefont {Pasquale}\ \bibnamefont {Maddaloni}},
  \bibinfo {author} {\bibfnamefont {Luigi}\ \bibnamefont {Santamaria}},
  \bibinfo {author} {\bibfnamefont {Paolo}\ \bibnamefont {De~Natale}}, \ and\
  \bibinfo {author} {\bibfnamefont {Maurizio}\ \bibnamefont {De~Rosa}},\
  }\bibfield  {title} {\enquote {\bibinfo {title} {Frequency comb generation in
  quadratic nonlinear media},}\ }\href {\doibase 10.1103/PhysRevA.91.063839}
  {\bibfield  {journal} {\bibinfo  {journal} {Phys. Rev. A}\ }\textbf {\bibinfo
  {volume} {91}},\ \bibinfo {pages} {063839} (\bibinfo {year}
  {2015})}\BibitemShut {NoStop}%
\bibitem [{\citenamefont {Leo}\ \emph {et~al.}(2016)\citenamefont {Leo},
  \citenamefont {Hansson}, \citenamefont {Ricciardi}, \citenamefont {De~Rosa},
  \citenamefont {Coen}, \citenamefont {Wabnitz},\ and\ \citenamefont
  {Erkintalo}}]{leo-PhysRevLett.116.033901}%
  \BibitemOpen
  \bibfield  {author} {\bibinfo {author} {\bibfnamefont {F.}~\bibnamefont
  {Leo}}, \bibinfo {author} {\bibfnamefont {T.}~\bibnamefont {Hansson}},
  \bibinfo {author} {\bibfnamefont {I.}~\bibnamefont {Ricciardi}}, \bibinfo
  {author} {\bibfnamefont {M.}~\bibnamefont {De~Rosa}}, \bibinfo {author}
  {\bibfnamefont {S.}~\bibnamefont {Coen}}, \bibinfo {author} {\bibfnamefont
  {S.}~\bibnamefont {Wabnitz}}, \ and\ \bibinfo {author} {\bibfnamefont
  {M.}~\bibnamefont {Erkintalo}},\ }\bibfield  {title} {\enquote {\bibinfo
  {title} {Walk-off-induced modulation instability, temporal pattern formation,
  and frequency comb generation in cavity-enhanced second-harmonic
  generation},}\ }\href {\doibase 10.1103/PhysRevLett.116.033901} {\bibfield
  {journal} {\bibinfo  {journal} {Phys. Rev. Lett.}\ }\textbf {\bibinfo
  {volume} {116}},\ \bibinfo {pages} {033901} (\bibinfo {year}
  {2016})}\BibitemShut {NoStop}%
\bibitem [{\citenamefont {Jung}\ \emph {et~al.}(2016)\citenamefont {Jung},
  \citenamefont {Guo}, \citenamefont {Zhu}, \citenamefont {Papp}, \citenamefont
  {Diddams},\ and\ \citenamefont {Tang}}]{Jung2016}%
  \BibitemOpen
  \bibfield  {author} {\bibinfo {author} {\bibfnamefont {Hojoong}\ \bibnamefont
  {Jung}}, \bibinfo {author} {\bibfnamefont {Xiang}\ \bibnamefont {Guo}},
  \bibinfo {author} {\bibfnamefont {Na}~\bibnamefont {Zhu}}, \bibinfo {author}
  {\bibfnamefont {Scott~B.}\ \bibnamefont {Papp}}, \bibinfo {author}
  {\bibfnamefont {Scott~A.}\ \bibnamefont {Diddams}}, \ and\ \bibinfo {author}
  {\bibfnamefont {Hong~X.}\ \bibnamefont {Tang}},\ }\bibfield  {title}
  {\enquote {\bibinfo {title} {Phase-dependent interference between frequency
  doubled comb lines in a $\chi^{(2)}$ phase-matched aluminum nitride
  microring},}\ }\href {\doibase 10.1364/OL.41.003747} {\bibfield  {journal}
  {\bibinfo  {journal} {Opt. Lett.}\ }\textbf {\bibinfo {volume} {41}},\
  \bibinfo {pages} {3747--3750} (\bibinfo {year} {2016})}\BibitemShut {NoStop}%
\bibitem [{\citenamefont {Hansson}\ \emph {et~al.}(2016)\citenamefont
  {Hansson}, \citenamefont {Leo}, \citenamefont {Erkintalo}, \citenamefont
  {Anthony}, \citenamefont {Coen}, \citenamefont {Ricciardi}, \citenamefont
  {Rosa},\ and\ \citenamefont {Wabnitz}}]{Hansson2016}%
  \BibitemOpen
  \bibfield  {author} {\bibinfo {author} {\bibfnamefont {Tobias}\ \bibnamefont
  {Hansson}}, \bibinfo {author} {\bibfnamefont {Fran\c{c}ois}\ \bibnamefont
  {Leo}}, \bibinfo {author} {\bibfnamefont {Miro}\ \bibnamefont {Erkintalo}},
  \bibinfo {author} {\bibfnamefont {Jessienta}\ \bibnamefont {Anthony}},
  \bibinfo {author} {\bibfnamefont {St\'{e}phane}\ \bibnamefont {Coen}},
  \bibinfo {author} {\bibfnamefont {Iolanda}\ \bibnamefont {Ricciardi}},
  \bibinfo {author} {\bibfnamefont {Maurizio~De}\ \bibnamefont {Rosa}}, \ and\
  \bibinfo {author} {\bibfnamefont {Stefan}\ \bibnamefont {Wabnitz}},\
  }\bibfield  {title} {\enquote {\bibinfo {title} {Single envelope equation
  modeling of multi-octave comb arrays in microresonators with quadratic and
  cubic nonlinearities},}\ }\href {\doibase 10.1364/JOSAB.33.001207} {\bibfield
   {journal} {\bibinfo  {journal} {J. Opt. Soc. Am. B}\ }\textbf {\bibinfo
  {volume} {33}},\ \bibinfo {pages} {1207--1215} (\bibinfo {year}
  {2016})}\BibitemShut {NoStop}%
\bibitem [{\citenamefont {Conforti}\ \emph
  {et~al.}(2010{\natexlab{b}})\citenamefont {Conforti}, \citenamefont
  {Baronio},\ and\ \citenamefont {De~Angelis}}]{conforti:2010}%
  \BibitemOpen
  \bibfield  {author} {\bibinfo {author} {\bibfnamefont {M.}~\bibnamefont
  {Conforti}}, \bibinfo {author} {\bibfnamefont {F.}~\bibnamefont {Baronio}}, \
  and\ \bibinfo {author} {\bibfnamefont {C.}~\bibnamefont {De~Angelis}},\
  }\bibfield  {title} {\enquote {\bibinfo {title} {Ultrabroadband optical
  phenomena in quadratic nonlinear media},}\ }\href {\doibase
  10.1109/JPHOT.2010.2051537} {\bibfield  {journal} {\bibinfo  {journal} {IEEE
  Photon. J.}\ }\textbf {\bibinfo {volume} {2}},\ \bibinfo {pages} {600--610}
  (\bibinfo {year} {2010}{\natexlab{b}})}\BibitemShut {NoStop}%
\bibitem [{\citenamefont {Guo}\ \emph {et~al.}(2013)\citenamefont {Guo},
  \citenamefont {Zeng}, \citenamefont {Zhou},\ and\ \citenamefont
  {Bache}}]{Guo:2013}%
  \BibitemOpen
  \bibfield  {author} {\bibinfo {author} {\bibfnamefont {Hairun}\ \bibnamefont
  {Guo}}, \bibinfo {author} {\bibfnamefont {Xianglong}\ \bibnamefont {Zeng}},
  \bibinfo {author} {\bibfnamefont {Binbin}\ \bibnamefont {Zhou}}, \ and\
  \bibinfo {author} {\bibfnamefont {Morten}\ \bibnamefont {Bache}},\ }\bibfield
   {title} {\enquote {\bibinfo {title} {Nonlinear wave equation in frequency
  domain: accurate modeling of ultrafast interaction in anisotropic nonlinear
  media},}\ }\href {\doibase 10.1364/JOSAB.30.000494} {\bibfield  {journal}
  {\bibinfo  {journal} {J. Opt. Soc. Am. B}\ }\textbf {\bibinfo {volume}
  {30}},\ \bibinfo {pages} {494--504} (\bibinfo {year} {2013})}\BibitemShut
  {NoStop}%
\bibitem [{\citenamefont {Gayer}\ \emph {et~al.}(2008)\citenamefont {Gayer},
  \citenamefont {Sacks}, \citenamefont {Galun},\ and\ \citenamefont
  {Arie}}]{gayer:2008}%
  \BibitemOpen
  \bibfield  {author} {\bibinfo {author} {\bibfnamefont {O.}~\bibnamefont
  {Gayer}}, \bibinfo {author} {\bibfnamefont {Z.}~\bibnamefont {Sacks}},
  \bibinfo {author} {\bibfnamefont {E.}~\bibnamefont {Galun}}, \ and\ \bibinfo
  {author} {\bibfnamefont {A.}~\bibnamefont {Arie}},\ }\bibfield  {title}
  {\enquote {\bibinfo {title} {Temperature and wavelength dependent refractive
  index equations for {M}g{O}-doped congruent and stoichiometric
  {L}i{N}b{O}$_3$},}\ }\href {\doibase 10.1007/s00340-008-2998-2} {\bibfield
  {journal} {\bibinfo  {journal} {Appl. Phys. B}\ }\textbf {\bibinfo {volume}
  {91}},\ \bibinfo {pages} {343--348} (\bibinfo {year} {2008})}\BibitemShut
  {NoStop}%
\bibitem [{\citenamefont {Lee}\ \emph {et~al.}(2000)\citenamefont {Lee},
  \citenamefont {Meade}, \citenamefont {Perlin}, \citenamefont {Winful},
  \citenamefont {Norris},\ and\ \citenamefont {Galvanauskas}}]{Lee2000}%
  \BibitemOpen
  \bibfield  {author} {\bibinfo {author} {\bibfnamefont {Y.-S.}\ \bibnamefont
  {Lee}}, \bibinfo {author} {\bibfnamefont {T.}~\bibnamefont {Meade}}, \bibinfo
  {author} {\bibfnamefont {V.}~\bibnamefont {Perlin}}, \bibinfo {author}
  {\bibfnamefont {H.}~\bibnamefont {Winful}}, \bibinfo {author} {\bibfnamefont
  {T.~B.}\ \bibnamefont {Norris}}, \ and\ \bibinfo {author} {\bibfnamefont
  {A.}~\bibnamefont {Galvanauskas}},\ }\bibfield  {title} {\enquote {\bibinfo
  {title} {Generation of narrow-band terahertz radiation via optical
  rectification of femtosecond pulses in periodically poled lithium niobate},}\
  }\href {\doibase http://dx.doi.org/10.1063/1.126390} {\bibfield  {journal}
  {\bibinfo  {journal} {Applied Physics Letters}\ }\textbf {\bibinfo {volume}
  {76}},\ \bibinfo {pages} {2505--2507} (\bibinfo {year} {2000})}\BibitemShut
  {NoStop}%
\bibitem [{\citenamefont {L'huillier}\ \emph
  {et~al.}(2007{\natexlab{a}})\citenamefont {L'huillier}, \citenamefont
  {Torosyan}, \citenamefont {Theuer}, \citenamefont {Avetisyan},\ and\
  \citenamefont {Beigang}}]{Lhuillier2007a}%
  \BibitemOpen
  \bibfield  {author} {\bibinfo {author} {\bibfnamefont {J.A.}\ \bibnamefont
  {L'huillier}}, \bibinfo {author} {\bibfnamefont {G.}~\bibnamefont
  {Torosyan}}, \bibinfo {author} {\bibfnamefont {M.}~\bibnamefont {Theuer}},
  \bibinfo {author} {\bibfnamefont {Y.}~\bibnamefont {Avetisyan}}, \ and\
  \bibinfo {author} {\bibfnamefont {R.}~\bibnamefont {Beigang}},\ }\bibfield
  {title} {\enquote {\bibinfo {title} {Generation of {THz} radiation using
  bulk, periodically and aperiodically poled lithium niobate -- part 1:
  Theory},}\ }\href {\doibase 10.1007/s00340-006-2490-9} {\bibfield  {journal}
  {\bibinfo  {journal} {Appl. Phys. B}\ }\textbf {\bibinfo {volume} {86}},\
  \bibinfo {pages} {185--196} (\bibinfo {year}
  {2007}{\natexlab{a}})}\BibitemShut {NoStop}%
\bibitem [{\citenamefont {L'huillier}\ \emph
  {et~al.}(2007{\natexlab{b}})\citenamefont {L'huillier}, \citenamefont
  {Torosyan}, \citenamefont {Theuer}, \citenamefont {Rau}, \citenamefont
  {Avetisyan},\ and\ \citenamefont {Beigang}}]{Lhuillier2007}%
  \BibitemOpen
  \bibfield  {author} {\bibinfo {author} {\bibfnamefont {J.A.}\ \bibnamefont
  {L'huillier}}, \bibinfo {author} {\bibfnamefont {G.}~\bibnamefont
  {Torosyan}}, \bibinfo {author} {\bibfnamefont {M.}~\bibnamefont {Theuer}},
  \bibinfo {author} {\bibfnamefont {C.}~\bibnamefont {Rau}}, \bibinfo {author}
  {\bibfnamefont {Y.}~\bibnamefont {Avetisyan}}, \ and\ \bibinfo {author}
  {\bibfnamefont {R.}~\bibnamefont {Beigang}},\ }\bibfield  {title} {\enquote
  {\bibinfo {title} {Generation of {THz} radiation using bulk, periodically and
  aperiodically poled lithium niobate -- part 2: Experiments},}\ }\href
  {\doibase 10.1007/s00340-006-2528-z} {\bibfield  {journal} {\bibinfo
  {journal} {Appl. Phys. B}\ }\textbf {\bibinfo {volume} {86}},\ \bibinfo
  {pages} {197--208} (\bibinfo {year} {2007}{\natexlab{b}})}\BibitemShut
  {NoStop}%
\bibitem [{\citenamefont {Shoji}\ \emph {et~al.}(1997)\citenamefont {Shoji},
  \citenamefont {Kondo}, \citenamefont {Kitamoto}, \citenamefont {Shirane},\
  and\ \citenamefont {Ito}}]{Shoji:1997}%
  \BibitemOpen
  \bibfield  {author} {\bibinfo {author} {\bibfnamefont {Ichiro}\ \bibnamefont
  {Shoji}}, \bibinfo {author} {\bibfnamefont {Takashi}\ \bibnamefont {Kondo}},
  \bibinfo {author} {\bibfnamefont {Ayako}\ \bibnamefont {Kitamoto}}, \bibinfo
  {author} {\bibfnamefont {Masayuki}\ \bibnamefont {Shirane}}, \ and\ \bibinfo
  {author} {\bibfnamefont {Ryoichi}\ \bibnamefont {Ito}},\ }\bibfield  {title}
  {\enquote {\bibinfo {title} {Absolute scale of second-order nonlinear-optical
  coefficients},}\ }\href {\doibase 10.1364/JOSAB.14.002268} {\bibfield
  {journal} {\bibinfo  {journal} {J. Opt. Soc. Am. B}\ }\textbf {\bibinfo
  {volume} {14}},\ \bibinfo {pages} {2268--2294} (\bibinfo {year}
  {1997})}\BibitemShut {NoStop}%
\end{thebibliography}

%

\end{document}